\documentclass[aps,prb,10pt,a4paper,twocolumn,footinbib, superscriptaddress]{revtex4-2}
\usepackage{graphicx}
\usepackage{amsmath}
\usepackage{amsfonts}
\usepackage{amssymb}
\usepackage{soul}
\usepackage{dsfont}
\usepackage{hyperref}
\usepackage{amstext}
\usepackage[caption=false]{subfig}
\usepackage{epstopdf} 
\usepackage{mathtools} 
\usepackage{balance}
\usepackage[caption=false]{subfig}

\hypersetup{colorlinks=true, citecolor=blue, urlcolor=blue, linkcolor=blue}
\usepackage{color,soul}
\usepackage{xcolor}
\usepackage{physics}
\usepackage{braket}
\usepackage{afterpage}
\usepackage{appendix}

\renewcommand{\rm}{\mathrm}

\date{\today}

\begin{document}

\title{Optimal quantum parametric feedback cooling}

\author{Sreenath K. Manikandan}
\email{sreenath.k.manikandan@su.se}
\affiliation{Nordita, KTH Royal Institute of Technology and Stockholm University, Hannes Alfv\'{e}ns v\"{a}g 12, SE-106 91 Stockholm, Sweden}
\author{Sofia Qvarfort}
\email{sofia.qvarfort@fysik.su.se}
\affiliation{Nordita, KTH Royal Institute of Technology and Stockholm University, Hannes Alfv\'{e}ns v\"{a}g 12, SE-106 91 Stockholm, Sweden}
\affiliation{Department of Physics, Stockholm University, AlbaNova University Center, SE-106 91 Stockholm, Sweden}

\begin{abstract}
We propose an optimal protocol using phase-preserving quantum measurements and phase-dependent modulations of the trapping potential at parametric resonance to cool a quantum oscillator to an occupation number of less than one quantum. We derive the optimal phase relationship and duration for the parametric modulations, and compute the lowest-possible occupation number in the steady state. The protocol is robust against moderate amounts of dissipation and phase errors in the feedback loop. Our work has implications for the cooling of levitated mechanical resonators in the quantum regime. 
\end{abstract}

\maketitle

\section{Introduction} Recent advances in fabricating and integrating devices in the nanoscale have made it possible to realize several candidate physical systems where quantum-mechanical behaviors are readily observed. Examples of this include superconducting quantum circuits~\cite{,wendin2017quantum,kjaergaard2020superconducting}, ultracold atoms~\cite{saffman2010quantum,tomza2019cold}, ion traps~\cite{cho2015review}, electron-spin qubits in semi-conductor platforms~\cite{burkard2021semiconductor}, and nanomechanical oscillators~\cite{aspelmeyer2014cavity}. 
 Notable achievements include the ability to prepare desired quantum-mechanical states on demand~\cite{hofheinz2008generation,matsukevich2004quantum} and perform gate operations~\cite{makhlin2001quantum,wendin2017quantum,reuer2022realization} as well as quantum-limited measurements~\cite{wiseman2009quantum} and real-time feedback control~\cite{vijay2012stabilizing,doherty1999feedback,doherty2000quantum}. 

Cooling has been one of the most significant challenges~\cite{giazotto2006opportunities}, an important example being cooling 
 mechanical oscillators in the quantum regime. These range from
 moving-end mirror Fabry--P\'{e}rot cavities and clamped membrane oscillators
 ~\cite{marquardt2009optomechanics,aspelmeyer2014cavity} to levitated systems~\cite{millen2020optomechanics,gonzalez2021levitodynamics}, quantum $LC$ circuits~\cite{vool2017introduction}, and hybrid optomechanical systems~\cite{rogers2014hybrid}. Achieving the quantum-mechanical ground state of oscillators via cooling is central to the exploration of various fundamental physics questions, such as  sensing weak forces and gravitational effects~\cite{qvarfort2018gravimetry,rademacher2020quantum, qvarfort2021optimal,moore2021searching,qvarfort2021constraining},  maintaining long-enough coherence times for information processing tasks~\cite{stannigel2012optomechanical}, and probing fundamental physics~\cite{ulbricht2021testing}.
Some of the experimentally implemented cooling protocols include
resolved-sideband cooling~\cite{teufel2011sideband,chan2011laser}, velocity damping~\cite{li2013millikelvin,tebbenjohanns2019cold}, Doppler cooling~\cite{barker2010doppler}, and coherent scattering~\cite{delic2020cooling}. In addition, a number of recent proposals have been put forward~\cite{montenegro2018ground,kounalakis2019synthesizing,zoepfl2022kerr}.   

\begin{figure}[t]
\includegraphics[width=0.6\linewidth]{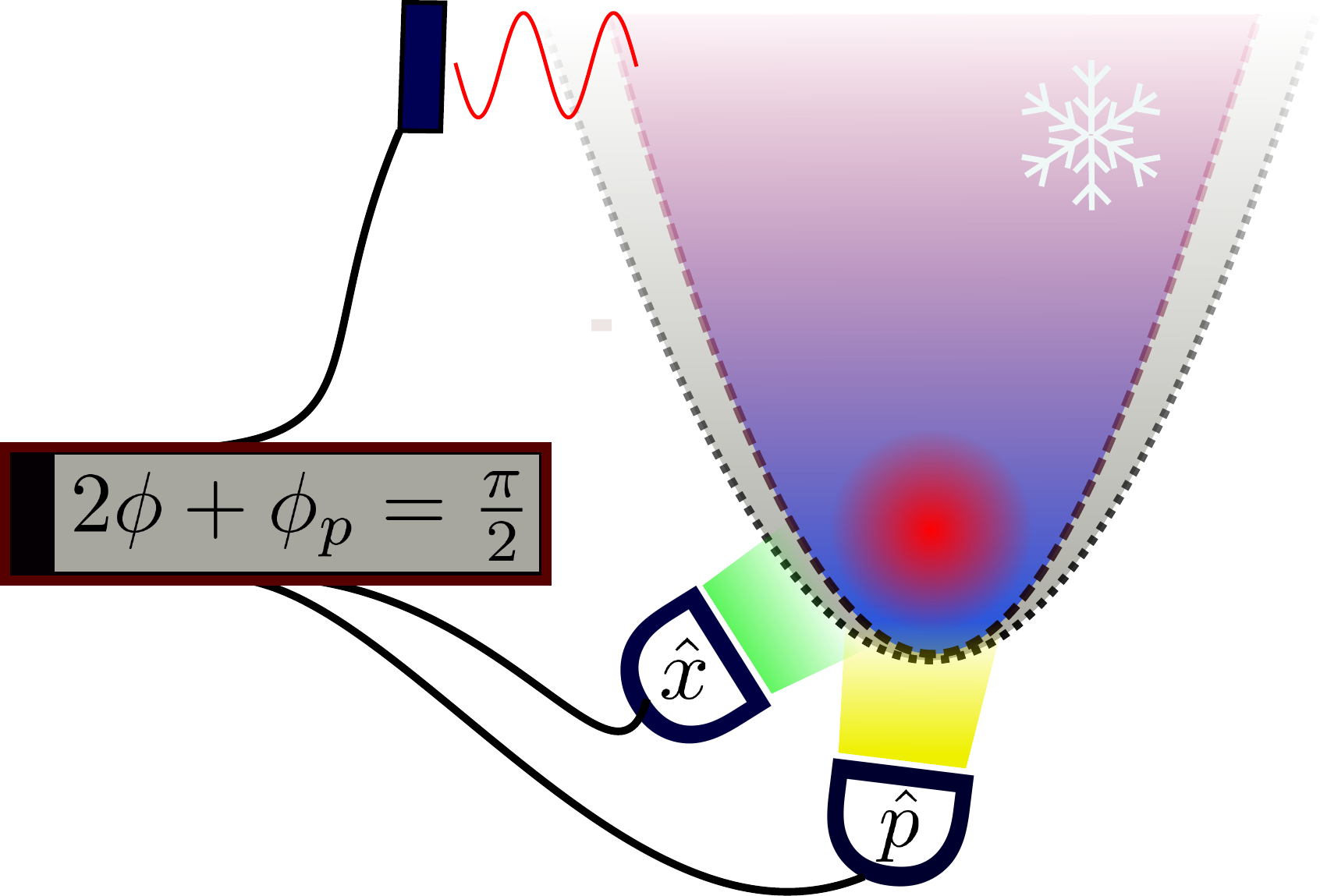}
\caption{Cooling a quantum particle in a harmonic trap by phase-preserving quantum measurements and phase-dependent parametric modulations of the trapping potential. In the feedback protocol, the phase information $\phi$ acquired from heterodyne measurements is used to update the phase $\phi_{p}$ of the parametric modulation of the trapping potential by requiring $2\phi+\phi_{p}=\frac{\pi}{2}$.
\label{figmain}}
\end{figure}

One such cooling method, known as parametric feedback cooling, has been especially successful in achieving ground-state cooling in levitated systems when used together with linear feedback techniques~\cite{gieseler2012subkelvin,jain2016direct, tebbenjohanns2021quantum,magrini2021real}.
 However, it is not clear whether parametric feedback cooling on its own can achieve ground-state cooling, or what its limits in the  quantum regime are. Motivated by this, the present work investigates the quantum regime of parametric feedback cooling of a simple harmonic oscillator. Henceforth, by ``cooling" we are referring to reducing the mean quanta in an oscillator. 
 Modulations of the harmonic potential at parametric resonance with a phase-offset are modeled by Mathieu’s equation~\cite{qvarfort2020time,qvarfort2021optimal}, which was first discussed by Mathieu~\cite{Mathieu1868}. 
 We show that parametric modulations with a definite phase reference relative to the oscillator state result in a reduction of  the mean quanta in the oscillator. In order to cool down arbitrary quantum states of the oscillator which lack a fixed phase reference (such as thermal states), we introduce phase-preserving (heterodyne) quantum measurements into the protocol (see Fig.~\ref{figmain}). 
 We then derive conditions for an optimal modulation time based on the measurement outcome and compute the steady-state occupation averaged number over many cooling cycles. We find it to be below one quantum, thus achieving near-quantum ground-state cooling.

This article is organized as follows. In Sec.~\ref{sec2dynamics} we summarize the methods for studying the parametrically driven dynamics of the oscillator. In Sec.~\ref{sec3cooling} we discuss the cooling protocol, and derive optimal driving phase and duration to achieve the near-quantum ground state by sequential cooling cycles which incorporate phase-preserving quantum measurements. We discuss the robustness to phase errors, and possible experimental implementations in Sec.~\ref{sec4discussions}. We conclude by discussing some of the future directions in Sec.~\ref{sec5conclusions}.

\section{Dynamics\label{sec2dynamics}}
In this section, we outline the solution of the quantum dynamics that results from the parametric modulations. In particular, we show how the quantum state of the oscillator undergoes single-mode squeezing.

\subsection{Solution to the dynamics} \label{sec:dynamics:solution}
The Hamiltonian describing the parametrically driven quantum oscillator has the form
\begin{equation}\label{eq:main:Hamiltonian1}
\hat H(t) = \hat H_0 +  2 m \omega_0 f(t) \hat x^2= \hat H_0 +  \hbar  f(t) \left( \hat a^\dag + \hat a \right)^2,
\end{equation}
where $\hat H_0 = \hbar \omega_{0} \left( \hat a^\dag \hat a + \frac{1}{2}\right)$
is the free Hamiltonian of the quantum oscillator, 
 $m$ is the mass of the oscillator, and $\omega_0$ is the frequency of the mode. In this work, we consider the following sinusoidal driving profile $f(t) = \lambda \cos(\omega_p t + \phi_p)$,
where $\lambda$ is the driving amplitude, $\omega_p$ is the drive frequency,  and $\phi_p$ is the phase. When $\omega_p = 2\omega_0$, the drive is referred to as parametric.
 
We describe the dynamics governed by the Hamiltonian in Eq.~\eqref{eq:main:Hamiltonian1} using the solutions developed in Refs.~\cite{qvarfort2020time,schneiter2020optimal,qvarfort2021optimal} (revisited in Appendix~\ref{app:relations}). 
In what follows, we briefly summarize the solutions. Since the Hamiltonian in Eq.~\eqref{eq:main:Hamiltonian1} is quadratic in its operator arguments, the evolution of a Gaussian state is  captured fully by 
the evolution of the first and second moments. Defining the vector of first moments as $\hat{\mathbb{X}} = (\hat a, \hat a^\dag )^{\mathrm{T}}$, the solution to the dynamics reads
\begin{equation}
\hat{\mathbb{X}}(t) = \hat U^\dag(t) \, \hat{\mathbb{X}} \, \hat U(t) \equiv \boldsymbol{S}(t) \, \hat{\mathbb{X}},
\end{equation}
where $\hat U(t)$ is the time-evolution operator given by 
\begin{align} \label{eq:time:evolution:operator}
\hat U(t) &= \overleftarrow{\mathcal{T}} \mathrm{exp} \left[ - \frac{i}{\hbar} \int^t_0 dt' \, \hat H(t') \right] 
\end{align}
and $\boldsymbol{S}(t)$ is a $2\times 2$ symplectic matrix given by 
\begin{align}
\boldsymbol{S}(t) &= \overleftarrow{\mathcal{T}} \mathrm{exp} \left[\boldsymbol{\Omega} \int^t_0 dt^\prime \, \boldsymbol{H}(t) \right],
\end{align}
for which  $\overleftarrow{\mathcal{T}}$ indicates time ordering of the exponential,  $\boldsymbol{\Omega}$ is the symplectic form, defined in this basis as $\boldsymbol{\Omega} = i \, \mathrm{diag}(-1,1)$, and $\boldsymbol{H}(t)$ is the Hamiltonian matrix, defined by $
\hat H(t) = \frac{1}{2}   \, \hat{\mathbb{X}}^\dag \, \boldsymbol{H}(t) \,  \hat{\mathbb{X}}. $
The $\boldsymbol{H}$ matrix corresponding to the Hamiltonian in Eq.~\eqref{eq:main:Hamiltonian1}  reads
\begin{align} \label{eq:Hamiltonian:matrix1}
\boldsymbol{H}(t) = \begin{pmatrix}  \hbar \omega_0 + 2  \hbar f(t)  & 2  \hbar f(t) \\ 2 \hbar f(t) &  \hbar \omega_0 + 2 \hbar f(t) \end{pmatrix}.  
\end{align}
The corresponding time evolution 
can be written as a Bogoliubov transformation of the first moments with
\begin{equation} \label{eq:Bogoliubov:S}
\boldsymbol{S}(t) = \begin{pmatrix} \alpha(t) & \beta(t) \\ \beta^*(t) & \alpha^*(t) \end{pmatrix}, 
\end{equation}
where $\alpha(t)$ and $\beta(t)$ are Bogoliubov coefficients 
satisfying 
$|\alpha(t)|^2 - |\beta(t)|^2 = 1$. 
The operator $\hat a(t)$ evolves as
\begin{align} 
\hat a(t) = \alpha(t) \, \hat a + \beta(t) \,\hat a^\dag.\label{eq:bg1}
\end{align}
The 
coefficients $\alpha(t)$ and $\beta(t)$ can be written as (see Appendix B in~\cite{qvarfort2020time})
\begin{equation} \label{eq:Bogoliubov:identities}
\begin{split}
\alpha(t) &= \frac{1}{2} \left( P(t)- i \, Q(t) +  \frac{d}{dt} \left[ i \,  P(t) + Q(t)\right] \right),  \\
\beta(t) &= \frac{1}{2} \left( P(t) + i \, Q(t) +  \frac{d}{dt} \left[ i \, P(t) - Q(t) \right] \right), 
\end{split}
\end{equation}
where the functions  $P(t)$ and $Q(t)$ are both solutions to the  differential equation
\begin{align} \label{eq:main:diff:eq1}
\ddot{y} + \left[ \omega_{0}^2 + 4\omega_{0} f(t) \right] y & = 0, 
\end{align}
where $P(t)$ can be obtained by using the initial conditions $P(0) =1$ and $\dot{P}(0) = 0$, and $Q(t)$ by setting $Q(0) = 0$ and $\dot{Q}(0) = 1$. These initial conditions follow from 
requiring that $\boldsymbol{S}(t = 0) = \mathds{1}$.  

The Hilbert space representation of this symplectic transformation is a rotation followed by a squeezing operation of the form $\hat S(z) = e^{(z \hat a^{\dag 2} - z^* \hat a^2)/2}$,
\begin{align} \label{eq:squeezing:expression}
\hat U(t) = e^{- i \varphi(t) \hat a^\dag \hat a } \, \hat S(z_{sq}(t)),   
\end{align}
where $\varphi(t) $ and $z_{sq} = r_{sq}(t) e^{i \theta_{sq}}(t)$ are time-dependent functions given in Appendix~\ref{app:relations}.

\subsection{Modulations at parametric resonance} 
When the frequency modulation occurs at twice the free frequency $\omega_p = 2\omega_{0}$,  Eq.~\eqref{eq:main:diff:eq1} takes the form of  Mathieu’s differential equation~\cite{Mathieu1868}
\begin{equation} \label{eq:mathieu}
\frac{d^2 y}{dx^2} + \bigl[ a - 2  \epsilon \cos(2 x  + \phi_p) \bigr] y = 0. 
\end{equation}
In our case $x = \omega_0 t$, $a = 1$, and $\epsilon = -2 \lambda/\omega_{0}$. Mathieu’s equation is usually defined without the phase  $\phi_p$. When $\phi_p = 0$, the solutions can be represented as Mathieu’s functions of the first kind: $\mathrm{ce}_n(t,\lambda)$ and $\mathrm{se}_n(t,\lambda)$. The solutions have no analytic form, but are periodic with $2\pi$.  We also note that at $a = 1$, the solutions are fundamentally unstable~\cite{bender1999advanced}. As a result, the system can only be operated at parametric resonance for short time-scales and with weak driving strengths $\lambda$.
A similar issue arises when considering the classical treatment of parametric feedback cooling~\cite{penny2021performance}. 
For non--zero $\phi_p$, the solutions can be expressed by linear combinations of $\mathrm{ce}_n(t, \lambda)$ and $\mathrm{se}_n(t,\lambda)$.  Mathieu's equation has been rigorously studied and describes the behavior of a diverse family of systems ranging from a child on a swing to the  buckling of membranes~\cite{budiansky1967dynamic}, as well as the quantum pendulum~\cite{condon1928physical}.

While Eq.~\eqref{eq:mathieu} does not allow for an exact analytical solution, an approximate solution can be obtained when $ \lambda/\omega_0 \ll 1$ using a two-time--scale method (see Ref~\cite{qvarfort2020time} and Appendix~\ref{app:dynamics}, which also includes a discussion of the errors in the approximation). Using this technique, the 
 Bogoliubov coefficients in Eq.~\eqref{eq:Bogoliubov:identities} can 
be approximated as, 
\begin{equation} \label{eq:approximate:Bogoliubov1}
\begin{split}
\alpha(t) &\approx e^{- i \omega_0 t} - i \frac{\lambda}{\omega_0} \, e^{i \phi_p} \sin(\omega_0 t),  \\
\beta(t) &\approx - i \lambda \, t \, e^{-i(\omega_0 t+ \phi_p)} . 
\end{split}
\end{equation}
We see that when $\lambda \rightarrow 0 $,  we are left with the free evolution $e^{- i \omega_0 t} $ encoded in $\alpha(t)$. 
 
\begin{figure*}
\includegraphics[width=0.7\linewidth]{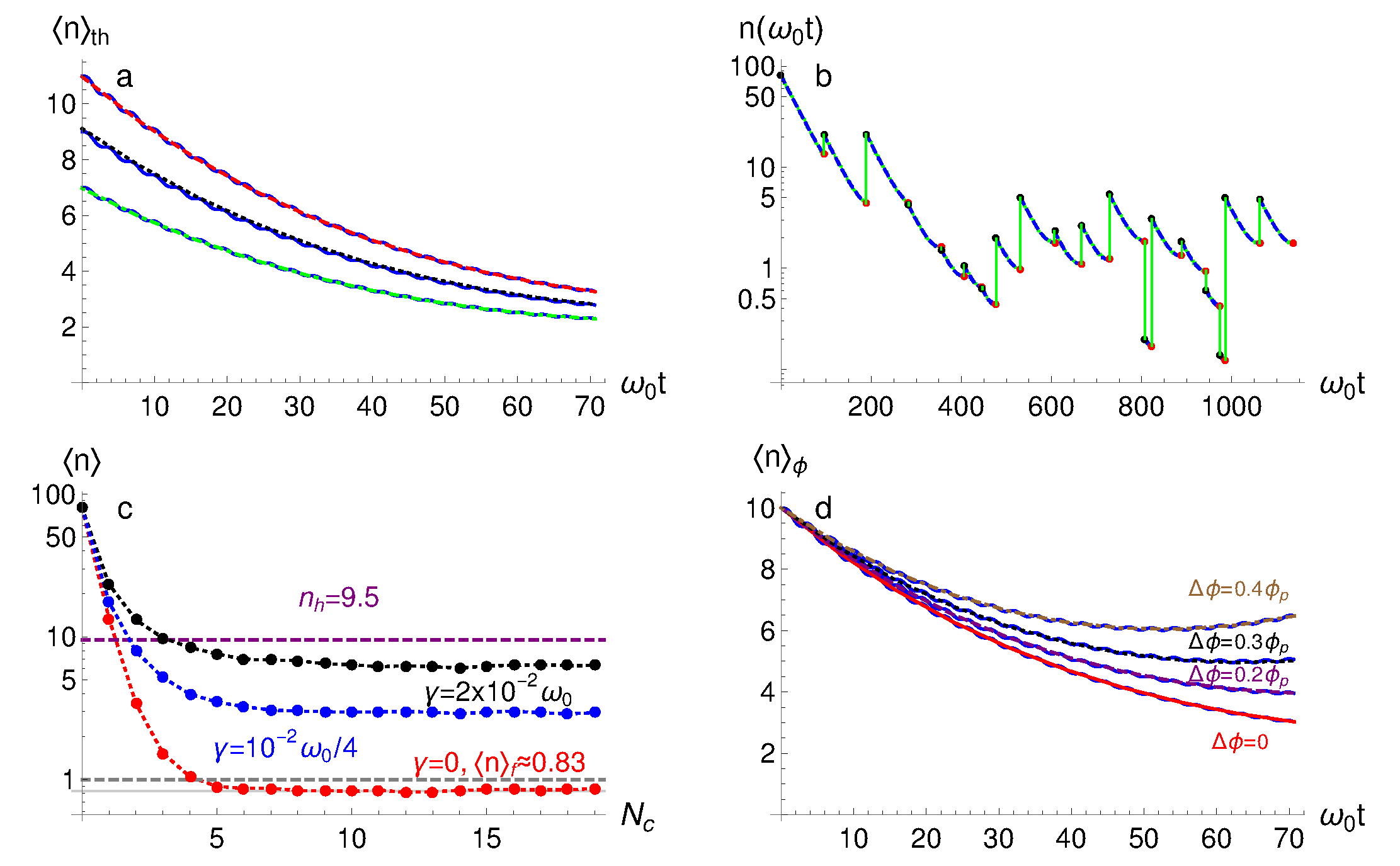}
\caption{ Results for optimal parametric feedback cooling.
(a) Average cooling of a thermal state of mean quanta $\bar{n}=10,~8$, and $6$. 
The  dashed lines indicate average cooling from numerical simulations of $10^{4}$ realizations of a single cooling cycle of duration $t = 45\pi/2\omega_{0}$, including the initial measurement that projects the state onto the coherent-state basis. The  blue solid curves indicate the corresponding analytical prediction. 
(b)  Single realization of the optimal cooling protocol  for an isolated system, starting from a coherent state with mean quanta equal to $r^2 = 80$.
The black dots indicate the average quanta in the coherent state obtained from measurements, and the red dots indicate the average quanta at the end of the cycle. The  blue dashed line indicates the change of mean quanta as the drive is turned on, and the underlying green line is the corresponding prediction from solving the dynamics analytically. 
(c) Average cooling $\braket{n}$  as a function of the number of cycles $N_c$  in the presence of a thermal environment, starting from a coherent state of  $r^2=80$. Here, $n_{h}$ is the thermal quantum of the ambient temperature, $k_{\text{B}}T_{B}=10\omega_{0}$.  In the absence of noise, the occupation number of the state can be successfully reduced to below one.
(d) Effects of uncertainly in phase control for different values of the phase spread $\Delta\phi$, in the absence of dissipation.  Each simulation starts with a coherent state with mean quanta, $r^2 = 10$. 
The blue curves indicate the corresponding theory predictions. All plots use the drive strength $\lambda/\omega_0 = 0.01$. \label{figplots}}
\end{figure*}

\section{Cooling through parametric modulations\label{sec3cooling}} 
In this section, we show that the dynamics that results from modulations at parametric resonance considered in Sec.~\ref{sec2dynamics} lead to cooling. We proceed to derive conditions for optimal cooling, which leads to the development of a cooling protocol. 

\subsection{Phase control for cooling}
We now derive the necessary phase relationship for cooling.  
Given Eq.~\eqref{eq:bg1}, we find that the mean occupation number in the oscillator changes as
\begin{equation}
\begin{split}
    \langle\hat{n}(t)\rangle & =\langle \hat{a}^{\dagger}(t)\hat{a}(t)\rangle = |\alpha(t)|^{2}\langle\hat{a}^{\dagger}\hat{a}\rangle+|\beta(t)|^{2}\langle\hat{a}\hat{a}^{\dagger}\rangle\\
&\quad + \alpha^{*}(t)\beta(t) \, \langle\hat{a}^{\dagger2}\rangle+\beta^{*}(t)\alpha(t) \, \langle\hat{a}^{2}\rangle.\label{fockn}
\end{split}
\end{equation}
Most experimental systems are initially found in thermal states, which lack a fixed phase reference. To obtain a phase relationship necessary for cooling, we instead start by considering the coherent-state basis. 
For an initial coherent state $|\xi\rangle=|re^{i\phi}\rangle$,  the mean occupation number evolves as
\begin{equation} \label{eq:coherent:state:cooling:exact}
\begin{split}
\braket{\hat n (t) }_{\ket{\xi}} & =  [|\alpha(t)|^2 + |\beta(t)|^2] | \, \xi|^2 + |\beta(t)|^2  \\
&\quad + \alpha^*(t) \,  \beta (t) \,  (\xi^*)^2 + \alpha(t) \, \beta^*(t) \,  \xi^2   .
\end{split}
\end{equation}
Using the approximate Bogoliubov coefficients in Eq.~\eqref{eq:approximate:Bogoliubov1}, the number of quanta evolves as, to first order in $\lambda$, 
\begin{equation} \label{eq:coherent:state:cooling1}
\begin{split}
\braket{\hat n (t)}_{\ket{\xi}} &\approx r^2 \biggl\{ 1  + \frac{\lambda}{\omega_0} \bigl[ \cos(\phi_p) - \cos(2 \omega_{0} t + \phi_p)  \\
&\qquad \qquad - 2 \omega_0 t \sin(2 \phi + \phi_p) \bigr] \biggr\}. 
\end{split}
\end{equation}
The last term inside the square brackets in Eq.~\eqref{eq:coherent:state:cooling1} is proportional to $\omega_0 t$, which means that it either increases or decreases the mean occupation number in the oscillator over time as determined by the phase relation $2 \phi + \phi_p$. When 
\begin{equation}
2\phi+\phi_p = \pi/2,\label{phaseCool}
\end{equation}
the last term in Eq.~\eqref{eq:coherent:state:cooling1} produces an initial cooling effect of rate $2 \lambda r^{2}.$\footnote{An analogous scenario occurs for the child in a swing problem, where a child crunches and stretches at twice the natural frequency of the swing  with the correct phase offset to increase or decrease the amplitude of the oscillations.} A proof that thermal states do not experience a reduction in their occupation number is presented in Appendix~\ref{thermalNote}.

\subsection{Optimal cooling}
In Sec.~\ref{sec:dynamics:solution}, we found that modulations at parametric resonance correspond to the single-mode squeezing operation $\hat S(z_{sq})$,  shown in Eq.~\eqref{eq:squeezing:expression}. However, such modulations cannot be used to  cool the initial coherent state indefinitely. For each initial coherent state, there exists an optimal squeezing value $r_{op}$ that maximally cools the coherent state. Squeezing beyond this value instead adds quanta to the state. 

For modulations at parametric resonance, the squeezing magnitude is given by $r_{sq}(t) \approx \lambda t$, where $t$ is the duration of the modulations (see Appendix~\ref{app:squeezing} for the derivation). Given the initial coherent state $\ket{\xi}$, we note that the optimal squeezing value is $r_{op}(t) = \ln( 1 + 4|\xi|^2)/4$. Thus, the optimal cooling time for a single cycle is 
\begin{align}
t_{op}\approx \mathrm{ln}(1 + 4|\xi|^2)/4 \lambda .
\end{align}
Furthermore, the minimum occupation number $n_{min}$ that can be achieved in each cycle starting from coherent state $|\xi\rangle$ is $n_{min}(\xi) = (\sqrt{1 + 4 |\xi|^2}- 1)/2$. Again see Appendix~\ref{app:squeezing} for detailed derivations.

\subsection{A single cooling cycle}
We now generalize the above result to a two-step  measurement-based feedback cooling protocol for cooling down arbitrary quantum initial states of the oscillator, including thermal states. The cooling cycle is described as follows. 
\begin{itemize}
    \item \textbf{Step 1.} Measure the quantum oscillator in the coherent-state basis
    ~\cite{arthurs_simultaneous_1965}. Let the measurement outcome be 
    $|\xi\rangle=|re^{i\phi}\rangle$.
    \item \textbf{Step 2.} Apply a conditional feedback modulation of the trapping potential  with phase offset $\phi_p = \pi/2 - 2\phi$ for the optimal duration $t_{op}\approx \mathrm{ln}(1 + 4 r^2)/4 \lambda $.
\end{itemize}

The measurements in the coherent-state basis, also known as heterodyne measurements, are required since parametric modulations alone cannot be used to decrease the mean quanta of thermal states. Furthermore, heterodyne measurements are optimal, since they only add a single quantum of noise on an average~\cite{manikandan2021efficiently} and because the phase-matching condition in Eq.~\eqref{phaseCool} that leads to cooling does not depend on the coherent-state amplitude (see Appendix~\ref{app:homodyne:measurements}). We demonstrate a single cycle for feedback cooling of a thermal state in Fig.~\ref{figplots}(a). A similar protocol with linear feedback has also been discussed as an engine in~\cite{manikandan2021efficiently}.

\subsection{Sequential cooling cycles}
We now consider applying a sequence of cooling cycles to a quantum oscillator, such that steps~1 and~2 above are repeated several times in sequence. For each cycle, the initial phase-preserving quantum measurements are modeled by Kraus operators, $K(\xi)=\frac{1}{\sqrt{\pi}}|\xi\rangle\langle\xi|$ sampling coherent states according to the corresponding Husimi $Q$ function at the beginning of each cycle. Since the dynamics is Gaussian, we approximate the sampling distribution as  (in units where $\hbar=m=1$)~\cite{husimi1940some,belenchia_entropy_2020,serafini2017quantum}
\begin{equation}
    Q(\text{Re}\xi,\text{Im}\xi)=\mathcal{MN}[\mu,\Sigma](\text{Re}\xi,\text{Im}\xi),
\end{equation}
where $\mathcal{MN}[\mu,\Sigma](\text{Re}\xi,\text{Im}\xi)$ is a multivariate normal distribution with mean 
\begin{align}
\mu=\begin{bmatrix}\sqrt{\omega_{0}/2}\langle \hat x\rangle\\\langle \hat p\rangle/\sqrt{2\omega_{0}}\end{bmatrix}, 
\end{align}
and a variance matrix defined as
\begin{equation}
    \Sigma = \frac{1}{2}\begin{bmatrix}\sigma_{xx}+1/2&\sigma_{xp}\\\sigma_{px}&\sigma_{pp}+1/2\end{bmatrix},
\end{equation}
where $\sigma_{ij}$ are the elements of the familiar covariance matrix of observables $\hat{i}$ and $\hat{j}$ defined as $\sigma_{ij}=\langle 1/2\{\hat{i}-\langle\hat{i}\rangle, \hat{j}-\langle\hat{j}\rangle\}\rangle$.  
In  the simulations, we assume that the measured coherent state is aligned to the $\phi=0$ axis via a unitary rotation such that $\phi_{p}=\pi/2$.  In each cycle, we compute the optimal cooling time $t_{op}$ as per the expressions above and truncate the modulations at the nearest multiple of~$\pi$.

In Fig.~\ref{figplots}(b) we show an example of such a quantum trajectory. The oscillator is initialized in a coherent state with mean quanta $r^2 = 80$,  and the optimal cycle duration is calculated for each subsequent measurement outcome. The occupation number $n(\omega_0t)$ quickly reduces to around unity.
An added benefit of performing the measurements is that the instability of the oscillator due to the Mathieu equation in the long-time limit is mitigated by the sequential measurements. 
Similar techniques have been employed in the past to improve the stability of otherwise unstable quantum oscillator systems~\cite{levy2016quantum}. 

\subsection{Near quantum ground-state cooling } \label{sec:near:ground:state}

 The optimized cooling scheme achieves cooling to near the quantum ground state. 
The probability of obtaining a state $\ket{\xi_j}$ from the $j$th measurement from the previous state $\ket{|\xi_{j-1}|}$ following modulations is 
\begin{align}
Q(\xi_j, |\xi_{j-1}|, r_{sq}) = \frac{1}{\pi} |\bra{\xi_{j}} \hat U(t) \ket{|\xi_{j-1}|}|^2, 
\end{align}
where $\hat U(t)$ is the time-evolution operator in Eq.~\eqref{eq:squeezing:expression}. 

As mentioned before, if the parametric modulations are turned on for the  optimal time $t_{op}$ starting from an initial coherent state $|\xi_j\rangle$, the achievable minimum occupation number in the oscillator at the end of the driving protocol is given by $ n_{min} (\xi_j) = (\sqrt{ 1 + 4 |\xi_j|^2} - 1)/2$. 
By then requiring that a single cycle does not, on average, change the occupation number of the system in the steady state, we find that $n_{min}(\xi_{j-1})|_{j\gg 1} \equiv \langle n\rangle_{f}$ also satisfies the requirement for an invariant cycle, namely,
\begin{equation}
n_{min} (\xi_{j-1})=\int d^{2}\xi_{j} \,  Q[\xi_j,|\xi_{j-1}|,r_{sq}(\xi_{j-1})] \, n_{min} (\xi_j), 
\end{equation}
which is a Fredholm integral equation of the first kind. We solve this equation numerically to prove that, in the absence of noise, the minimum occupation in the steady state is $\langle n\rangle_{f} \approx 0.83$ (see Appendix~\ref{app:trajectories}). This constitutes the optimal limit to parametric feedback cooling in the quantum regime in our protocol.

\vspace{1cm} 

\subsection{Cooling in a thermal environment} 
All experimental systems are affected by environmental noise.
We therefore consider performing multiple cycles of cooling while the quantum oscillator is undergoing collisional interactions with modes of a thermal reservoir. Following~\cite{leitch2022driven}, we model these interactions using an adiabatic Markovian master equation, resulting in the dynamical equations for the first and second moments
\begin{eqnarray}
\dot{x}(t)&=&p(t)-\gamma x(t)/2,~~\dot{p}(t)=-\omega(t)^{2}x(t)-\gamma p(t)/2,\nonumber\\
\dot{\sigma}_{xx}(t) &=&-\gamma \sigma_{xx}(t)+\gamma\frac{2\bar{n}_B+1}{2\omega(t)}+2\sigma_{xp}(t),\nonumber\\
\dot{\sigma}_{xp}(t)&=&-\gamma\sigma_{xp}(t)+\sigma_{pp}(t)-\sigma_{xx}\omega(t)^{2},~~\sigma_{xp}(t)=\sigma_{px}(t),\nonumber\\
\dot{\sigma}_{pp}(t)&=&-\gamma\sigma_{pp}(t)+\gamma\frac{(2\bar{n}_B+1)\omega(t)}{2}-2\sigma_{xp}\omega(t)^{2}, 
\end{eqnarray}
where $\gamma$ is the dissipation rate, $\bar{n}_B$ is the thermal occupation of a reservoir mode at frequency $\omega_{B}=\omega_{0}/2$, having a temperature $k_{\text{B}}T_{B}=10\omega_{0}$, and $\omega(t) = \omega_0 \sqrt{1 + 4 f(t)/\omega_0}$. We compute the optimal cooling time $t_{op}$ numerically for each cycle by searching for the minimum of the occupation value. 

In Fig.~\ref{figplots}(c), we demonstrate that our protocol is able to cool down the quantum oscillator below the ambient temperature on average, even for moderate amounts of dissipation. Here, we start from a coherent state at $r^2 = 80$ and average the result from $10^3$ runs.  For negligible dissipation, 
we recover the occupation value of 0.83 derived in Sec.~\ref{sec:near:ground:state}, which corresponds to cooling near the quantum ground state.

\section{Discussion\label{sec4discussions}}
Here we discuss the effects of phase noise on the cooling power, as well as the physical implementations of the protocol across different platforms. 

\subsection{Phase inaccuracy} The protocol relies on the ability to adjust the phase of the modulation to $2 \phi+\phi_p = \pi/2$. However, latency in the feedback loop and other inaccuracies can introduce errors into the protocol. To model this scenario, we consider several realizations of an individual cooling cycle where the driving phase $\phi_{p}^\prime$ is sampled around the ideal driving phase $\phi_p$ according to the probability distribution  
\begin{equation} 
P(\phi_{p}^\prime)= \frac{1}{\Delta\phi\sqrt{2\pi}}\mathrm{exp}\bigl[ -  ( \phi _p^\prime - \phi_p)^2/2\Delta \phi^2\bigr],
\end{equation}
with standard deviation  $\Delta \phi$. In Fig.~\ref{figplots}(d), we demonstrate that our cooling protocol is robust against significant phase errors up to $20\%$ of the ideal phase.

\subsection{Physical realization}
Modulations of the trapping potential can be realized by imposing an electrostatic force or external strong optical field on the mechanical mode~\cite{blencowe2004quantum}.  In levitated systems 
the percentage change of the trapping potential is known as the modulation depth $G$~\cite{penny2021performance}. In this work, $G$ is related to the driving amplitude $\lambda$ as $G = 4 \lambda/\omega_0$, which for $\lambda/\omega_0 = 0.01$ is $G = 0.04$ or $4\%$. In hybrid traps, modulation depths as high as $5\%$ are possible~\cite{penny2021performance}, while in optical tweezers, around $0.4\%$ is more common~\cite{vovrosh2017parametric}. Beyond optical and hybrid traps, candidate systems include magnetically levitated magnets~\cite{wang2019dynamics, vinante2020ultralow}, diamagnets~\cite{lewandowski2021high}, and superconducting spheres~\cite{latorre2021chip}. 

Phase-preserving measurements are a key ingredient in this protocol and can be implemented through joint homodyne detection of both quadratures of the oscillator~\cite{arthurs_simultaneous_1965,karmakar2022stochastic}, or by 
pulsing light through the cavity when the system is in the unresolved sideband regime~\cite{vanner2011pulsed, kanari2021two}. 
Superconducting 
circuits also offer novel methodologies to perform 
such measurements dynamically  in hybrid systems~\cite{campagne2016observing, karmakar2022stochastic}.

\section{Conclusions\label{sec5conclusions}} 
The optimal parametric feedback protocol proposed here leads to near quantum ground-state cooling, and appears to offer significant cooling even when feedback capabilities are limited. The protocol may also be combined with linear feedback cooling techniques~\cite{tebbenjohanns2021quantum,magrini2021real} or various other quantum refrigerator schemes proposed based on fundamental thermodynamic principles~\cite{karimi2016otto,levy2012quantum,manikandan2019superconducting,manikandan2020autonomous,fornieri20170}, 
to further explore quantum enhanced cooling at the nanoscale. The methodologies we developed can be generalized 
to derive exact results for optimal cycles in the presence of added noise; we defer this analysis to future work.

\textit{Note added}: Recently, the authors became aware of a related paper by Ghosh \textit{et al}.~\cite{ghosh2022theory}, where phase-adaptive quantum parametric feedback cooling is considered using a semi-classical approach. With the assumption of the equipartition of noise between the phase-space quadratures, the authors of~\cite{ghosh2022theory} demonstrate efficient, exponential cooling by deriving the same phase-relation as that found here.  In contrast, the present manuscript also highlights the role of squeezing that results from the parametric modulations of the trapping potential, which suggests an optimal duration of the cooling cycle. 

\section*{Acknowledgments} 
We thank Anthony Bonfils for helpful insights concerning the stability of Mathieu's equation, and added insights on the child in a swing problem. We also thank Lydia Kanari-Naish, Thomas Penny, Antonio Pontin, Anis Rahman, Ermes Scarano, Dhrubaditya Mitra, David Edward Bruschi, Alessio Serafini, and Witlef Wieczorek for helpful comments and discussions.  The work of S.K.M.~was supported by the Wallenberg Initiative on Networks and Quantum Information (WINQ). S.Q.~was funded in part by the Wallenberg Initiative on Networks and Quantum Information (WINQ) and in part by the Marie Skłodowska--Curie Action IF programme Nonlinear optomechanics for verification, utility, and sensing -- Grant-No.~101027183. Nordita is partially supported by Nordforsk.

\section*{Data availability statement}
The code used to generate the figures shown in this work can be found in \href{https://github.com/sqvarfort/parametric-feedback-cooling}{GitHub repository}.

\bibliographystyle{apsrev4-2}
\bibliography{refs}

\onecolumngrid

\appendix
\widetext

\section{Derivation of the dynamics} \label{app:relations}
In this appendix,  we connect the derivation of the solutions for the dynamics, which was first presented in~\cite{qvarfort2020time}, with a more intuitive solution using a Lie algebra method~\cite{wei1963lie} (see~\cite{qvarfort2022solving} for a pedagogical introduction). We identify a set of operators that is closed under commutation, which allows us to set up differential equations that, when solved, provide the exact solution to the dynamics. We then show that these solutions can be mapped to those derived in~\cite{qvarfort2020time}.  The solutions and the derivations build on methods also developed in Ref~\cite{bruschi2013time}. In addition, we note that the dynamics of this form may also be treated using the exact Lewis--Riesenfeld solutions~\cite{lewis1969exact}.

\subsection{Phase space dynamics}
We start by setting $\hbar = 1$ in this section. Then, we identify the elements of the Lie algebra that generate the time evolution induced by the Hamiltonian in Eq.~\eqref{eq:main:Hamiltonian1}. The elements are
\begin{align}
&\hat a^\dag \hat a, && \hat a^{\dag2} + \hat a^2,  && i \left( \hat a^{\dag 2} - \hat a^2 \right). 
\end{align}
It can be verified that the algebra is closed under commutation. 
The corresponding symplectic matrices in the $(\hat a, \hat a^\dag)^{\mathrm{T}}$ basis, which we call $\boldsymbol{A}_0$, $\boldsymbol{A}_+$, and $\boldsymbol{A}_-$, are given by 
\begin{align}
&\boldsymbol{A}_0 = \begin{pmatrix} 1 & 0 \\ 0 & 1 \end{pmatrix}, && \boldsymbol{A}_+ =2 \begin{pmatrix} 0 & 1 \\ 1 & 0 \end{pmatrix},  && \boldsymbol{A}_- = 2i  \begin{pmatrix} 0 & 1 \\ -1 & 0 \end{pmatrix}. 
\end{align}
The symplectic matrix that encodes the evolution of the system is given by 
\begin{align}
\boldsymbol{S}(t) = \mathcal{T}\mathrm{exp}\left[ \boldsymbol{\Omega} \int^t_0 dt^\prime \, \boldsymbol{H}(t^\prime) \right].
\end{align}
We then differentiate this matrix with respect to time $t$ to find 
\begin{equation} \label{app:eq:diff:of:S}
\frac{d}{dt} \boldsymbol{S}(t)  = \boldsymbol{\Omega} \boldsymbol{H}(t) \, \boldsymbol{S}(t). 
\end{equation}
We then multiply the expression by $\boldsymbol{S}^{-1}(t)$ on the right-hand side to find
\begin{equation}
\dot{\boldsymbol{S}}(t) \boldsymbol{S}^{-1}(t) = \boldsymbol{\Omega} \boldsymbol{H}(t) . 
\end{equation}
Then, we make the following ansatz for the solution to $\boldsymbol{S}(t)$:
\begin{equation} \label{eq:phase:space:ansatz}
\boldsymbol{S}(t) = e^{J_0 \, \boldsymbol{\Omega}  \boldsymbol{A}_0}\, e^{ J_+ \boldsymbol{\Omega} \boldsymbol{A}_+ } \, e^{J_- \boldsymbol{\Omega}\boldsymbol{A}_-}. 
\end{equation}
Here, $J_0$, $J_+$, and $J_-$ are time-dependent coefficients that we wish to find. 
We then differentiate the ansatz in Eq.~\eqref{eq:phase:space:ansatz} to find 
\begin{equation}
\dot{\boldsymbol{S}}(t) = \dot{J}_0 \boldsymbol{\Omega} \boldsymbol{A}_0 e^{J_0 \, \boldsymbol{\Omega}  \boldsymbol{A}_0}\, e^{ J_+ \boldsymbol{\Omega} \boldsymbol{A}_+ } \, e^{J_- \boldsymbol{\Omega}\boldsymbol{A}_-} \, + \dot{J}_+ e^{J_0 \, \boldsymbol{\Omega}  \boldsymbol{A}_0}\, \boldsymbol{\Omega}\boldsymbol{A}_+ e^{ J_+ \boldsymbol{\Omega} \boldsymbol{A}_+ } \, e^{J_- \boldsymbol{\Omega}\boldsymbol{A}_-} + \dot{J}_-e^{J_0 \, \boldsymbol{\Omega}  \boldsymbol{A}_0}\,e^{ J_+ \boldsymbol{\Omega} \boldsymbol{A}_+ } \,   \boldsymbol{\Omega}\boldsymbol{A}_- e^{J_- \boldsymbol{\Omega}\boldsymbol{A}_-}. 
\end{equation}
Multiplying by $\boldsymbol{S}^{-1}(t)$ on the right, we find 
\begin{equation}
\dot{\boldsymbol{S}}(t) \boldsymbol{S}^{-1}(t) = \dot{J}_0 \boldsymbol{\Omega} \boldsymbol{A}_0  + \dot{J}_+ e^{J_0 \, \boldsymbol{\Omega}  \boldsymbol{A}_0}\, \boldsymbol{\Omega}\boldsymbol{A}_+ e^{- J_+ \boldsymbol{\Omega}\boldsymbol{A}_0}  + \dot{J}_-e^{J_0 \, \boldsymbol{\Omega}  \boldsymbol{A}_0}\,e^{ J_+ \boldsymbol{\Omega} \boldsymbol{A}_+ } \,   \boldsymbol{\Omega}\boldsymbol{A}_-  \, e^{- J_+ \boldsymbol{\Omega} \boldsymbol{A}_+ } \, e^{-J_0 \boldsymbol{\Omega}\boldsymbol{A}_0}. 
\end{equation}
Multiplying both expressions by $\boldsymbol{\Omega}^{-1}$ on the left and using Eq.~\eqref{app:eq:diff:of:S} gives us 
\begin{equation} \label{app:eq:Hamiltonian:mid:step}
 \boldsymbol{H}(t)  = \dot{J}_0  \boldsymbol{A}_0  + \boldsymbol{\Omega}^{-1}\dot{J}_+ e^{J_0 \, \boldsymbol{\Omega} \boldsymbol{A}_0}\,\boldsymbol{\Omega}\boldsymbol{A}_+ e^{- J_+ \boldsymbol{\Omega}\boldsymbol{A}_0}  +\boldsymbol{\Omega}^{-1} \dot{J}_-e^{J_0 \, \boldsymbol{\Omega}  \boldsymbol{A}_0}\,e^{ J_+ \boldsymbol{\Omega} \boldsymbol{A}_+ } \, \boldsymbol{\Omega}\boldsymbol{A}_- e^{- J_+ \boldsymbol{\Omega} \boldsymbol{A}_+ } \, e^{-J_0 \boldsymbol{\Omega}\boldsymbol{A}_0}. 
\end{equation}
Then we also know that the symplectic matrices obey $\boldsymbol{S} \boldsymbol{\Omega} \boldsymbol{S}^\dag = \boldsymbol{\Omega}$. This allows us to rewrite Eq.~\eqref{app:eq:Hamiltonian:mid:step} as
\begin{equation}
 \boldsymbol{H}(t)  = \dot{J}_0  \boldsymbol{A}_0  + \dot{J}_+\,e^{J_0 \, \boldsymbol{\Omega} \boldsymbol{A}_0} \boldsymbol{A}_+ e^{- J_+ \boldsymbol{\Omega}\boldsymbol{A}_0}  +  \dot{J}_-  e^{J_0 \, \boldsymbol{\Omega}  \boldsymbol{A}_0}\,e^{ J_+ \boldsymbol{\Omega} \boldsymbol{A}_+ }\boldsymbol{A}_- e^{- J_+ \boldsymbol{\Omega} \boldsymbol{A}_+ } \, e^{-J_0 \boldsymbol{\Omega}\boldsymbol{A}_0}, 
\end{equation}
which, after multiplying out the matrices, leaves us with
\begin{equation} \label{app:eq:Hamiltoniana:diff:eq:coeffcs}
    \boldsymbol{H}(t) =  \begin{pmatrix} \dot{J}_0 + 2 \sinh(4 J_+) \dot{J}_- & 2 \, e^{- 2 i J_0} \left[ i \cosh(4 J_+) \dot{J}_- + \dot{J}_+ \right] \\
    e^{ 2 i J_0} \left[ - i \cosh(4 J_+) \dot{J}_- + \dot{J}_+ \right] &\dot{J}_0 + 2 \sinh(4 J_+) \dot{J}_- \end{pmatrix}.  
\end{equation}
However, we also know that the Hamiltonian matrix is given by 
\begin{align} \label{app:eq:Hamiltonian:matrix}
\boldsymbol{H}(t) =  \begin{pmatrix}  \omega_0 + 2  f(t)  & 2  f(t) \\
2 f(t) &  \omega_0 + 2 f(t) \end{pmatrix}. 
\end{align}
Equating Eqs.~\eqref{app:eq:Hamiltoniana:diff:eq:coeffcs} and~\eqref{app:eq:Hamiltonian:matrix} allows us to identify the differential equations
\begin{equation} \label{eq:diff:eqs}
\begin{split}
\omega_0 + 2 f(t) &= \dot{J}_0  + 2 \sinh(4 J_+) \dot{J}_-,   \\
2 f(t) &= 2 \, e^{- 2 i J_0} \left[ i \cosh(4 J_+) \dot{J}_- + \dot{J}_+\right]. 
\end{split}
\end{equation}
By manipulating the expressions in Eq.~\eqref{eq:diff:eqs}, it is possible to isolate the three coefficients $J_0$, $J_+$, and $J_-$ into the three differential equations~\cite{schneiter2020optimal}
\begin{align} \label{eq:diff:equations:Js}
\dot{J}_0 &=  \omega_0 + 2\,  f(t) \,  \left[ 1 - \sin(2 J_0) \tanh(4 J_+) \right],  \nonumber \\
\dot{J}_+ &=   f(t) \,  \cos(2 J_0)  ,  \\
\dot{J}_-  &=  f(t)\,  \frac{\sin(2 J_0)}{\cosh(4 J_+)} \nonumber .
\end{align}
We note, however, that ${J}_-$ does not feature in the first and second equations for $\dot{J}_0$ and $\dot{J}_+$, which means that it can be entirely solved once the other two equations have been solved. This confirms that $\boldsymbol{S}(t)$ is fully determined by only two real parameters. 

We now wish to relate $J_0$, $J_+$, and $J_-$ to the functions $P(t)$ and $Q(t)$, which were introduced in Eq.~\eqref{eq:Bogoliubov:identities}. For the derivation of $P(t)$ and $Q(t)$, see Appendix B in Ref~\cite{qvarfort2020time}. By rewriting $\boldsymbol{S}(t)$ in Eq.~\eqref{eq:phase:space:ansatz} as a single symplectic operator, we find that  the Bogoliubov coefficients $\alpha(t)$ and $\beta(t)$ can be written as~\cite{schneiter2020optimal}
\begin{equation} \label{eq:alpha:beta:from:S}
\begin{split}
\alpha(t) &= e^{- i J_0} \left[ \cosh(2 J_+)\cosh(2 J_-) - i \sinh(2 J_+) \sinh(2 J_-) \right],  \\
\beta(t) &= e^{- i J_0} \left[ \cosh(2 J_+) \sinh(2 J_-) - i\sinh(2 J_+) \cosh(2 J_-)  \right], 
\end{split}
\end{equation}
 where $|\alpha(t)|^2 - |\beta(t)|^2 = 1$.
Also from using Eq.~\eqref{eq:Bogoliubov:identities}, we are able to identify the relationships
\begin{align}
&P(t) = \mathrm{Re}[\alpha] + \mathrm{Re}[\beta], 
&&Q(t) = \mathrm{Im}[\beta] - \mathrm{Im}[\alpha], \nonumber \\
&\ddot{P}(t) = \mathrm{Im}[\dot{\alpha}] + \mathrm{Im}[\dot{\beta}],
&&\ddot{Q}(t) = \mathrm{Re}[\dot{\alpha}] - \mathrm{Re}[\dot{\beta}].
\end{align}
It is then possible to write $P(t)$ and $Q(t)$ in terms of $J_0, $ $J_+$, and $J_-$  as
\begin{equation}\label{app:eq:P:Q:to:Js}
\begin{split}
P(t) &=  e^{2 J_-} \left[ \cos(J_0) \cosh(2 J_+) - \sin(J_0) \sinh(2 J_+) \right],  \\
Q(t) &= e^{-2 J_- } \left[ \sin (J_0) \cosh (2 J_+ )-\cos (J_0) \sinh (2 J_+) \right].
\end{split}
\end{equation}
Similarly, the second derivatives $\ddot{P}(t)$ and $\ddot{Q}(t)$ can be found, which are long expressions, so we do not print them here.  
We then recall from the main text that $P(t)$ and $Q(t)$ are determined by the two differential equations
\begin{align} \label{app:eq:diff:eqs}
&\ddot{P}(t) + \bigl[1 + 4 f(t) /\omega_0 \bigr] P(t) = 0, 
&&\ddot{Q}(t) + \bigl[ 1 + 4 f(t)/\omega_0 \bigr] Q(t) = 0. 
\end{align}
By then inserting the expressions in Eq.~\eqref{app:eq:P:Q:to:Js} and their derivatives into Eq.~\eqref{app:eq:diff:eqs}, and using the relations in Eq.~\eqref{eq:diff:equations:Js}, it is possible to show that $J_0$, $J_+$ and $J_-$ and their relationship are also solutions to these equations. 

Next, we note that it is also possible to define $J_0$, $J_+$, and $J_-$ in terms of $P(t)$ and $Q(t)$. Previously, it was shown that~\cite{schneiter2020optimal}
\begin{align} \label{app:eq:Js:alpha:beta}
    \cosh(4 J_+) &= |\alpha^2(t) - \beta^2(t)|, \nonumber \\
    \cosh(4 J_-) &= \frac{|\alpha(t)|^2 + |\beta(t)|^2 }{|\alpha^2(t) - \beta^2(t)|},   \\
    e^{- 2 i J_0} &= \frac{\alpha^2(t) - \beta^2(t) }{|\alpha^2(t) - \beta^2(t)|}. \nonumber
\end{align}
With the help of the relations in Eq.~\eqref{app:eq:Js:alpha:beta}, we can identify 
\begin{align} \label{app:eq:relate:P:Q:J}
    \cosh(4 J_+) &= \Bigl|\left[ i \dot{P}(t) + P(t) \right]\left[ \dot{Q}(t) - i Q(t)\right]\Bigr|,  \nonumber \\
    \cosh(4 J_-) &=  \frac{1}{2} \frac{P^2(t) + Q^2(t) + \dot{P}^2(t) + \dot{Q}^2(t) }{\Bigl|\left[ i \dot{P}(t) + P(t) \right]\left[ \dot{Q}(t) - i Q(t)\right]\Bigr| },   \\
    e^{- 2 i J_0} &= 2\frac{\left[ i \dot{P}(t) + P(t) \right]\left[ \dot{Q}(t) - i Q(t)\right]}{\Bigl|P^2(t) + Q^2(t) + \dot{P}^2(t) + \dot{Q}^2(t) \Bigr|}. \nonumber
\end{align}
Finally, we note that the solutions $P(t)$ and $Q(t)$ are valid for \textit{any} choice of driving function $f(t)$. The case of parametric modulations explored in the main text leads to Mathieu's equation, but many other driving patterns can be considered using these methods.

\subsection{Hilbert space solution} 
Here, we use the solutions derived in the preceding section to cast the dynamics into a rotation and a single squeezing operator in the Hilbert space representation. 
The  Hilbert space ansatz equivalent to that in Eq.~\eqref{eq:phase:space:ansatz} is 
\begin{align} \label{eq:Hilbert:space:ansatz}
\hat U(t) = e^{- i J_0 \, \hat a^\dag \hat a} \, e^{- i J_+ \, (\hat a^{\dag 2} + \hat a^2 ) } \, e^{- i J_- [i (\hat a^{\dag 2} - \hat a^2 )]}. 
\end{align}
We note that the operators in Eq.~\eqref{eq:Hilbert:space:ansatz} are equivalent to single-mode squeezing and a phase rotation with $\hat a^\dag \hat a$. The connection between the Hilbert space picture and the phase-space picture is
\begin{align}
\hat U^\dag (t) \, \hat{\mathbb{X}} \, \hat U(t) &= \boldsymbol{S}(t) \, \hat{\mathbb{X}}, 
\end{align}
where $\hat{\mathbb{X}} = (\hat a, \hat a^\dag)^{\mathrm{T}}$, as in the main text. 

We start by focusing on the two squeezing operators $e^{-i J_+ (\hat a^{\dag 2} + \hat a^2)}$ and $e^{ J_-(\hat a^{\dag 2} -\hat a^2)} $. It is possible to combine two squeezing operators by using the product theorem~\cite{agarwal2012quantum}
\begin{align}
\hat S(z_1) \, \hat S(z_2) = e^{\varphi_{sq} ( \hat a^\dag \hat a + \hat a \hat a^\dag )} \hat S(z_3) , 
\end{align}
where the squeezing operators are defined as $S(z_j) = e^{(z_j^* \hat a ^2 - z_j \hat a^{\dag 2})/2}$ for  $z_j = r_j e^{i \theta_j}$. The phase $\varphi_{sq}$ is given by 
\begin{align}
\varphi_{sq} = \frac{1}{4} \ln\left( \frac{1 + t_1 t_2^*}{1  + t_1^* t_2} \right) , 
\end{align}
for which $t_j = \tanh(r_j) e^{i \theta_j}$. It then follows that 
\begin{align}
t_3 = \tanh(r_3) e^{i \theta_3} = \frac{t_1 + t_2}{1 + t_1 t_2^*} .
\end{align}
We wish to solve for the total squeezing value $r_{3}$ and determine its behavior given the parametric modulations. By  making the identification that in our case, we have 
\begin{align}
r_1 &= 2J_+, && \theta_1 = \pi/2, \nonumber \\
r_2 &= 2J_-, && \theta_2 = \pi, 
\end{align}
we find~\cite{schneiter2020optimal}
\begin{align} \label{app:eq:tanhr3}
\tanh(r_3) e^{i \theta_3} = \frac{i\tanh(2J_+) - \tanh(2J_-)}{1-i\tanh(2J_+)\tanh(2J_-)}.
\end{align}
To find an expression for $\tanh(r_3)$, we take the absolute value of Eq.~\eqref{app:eq:tanhr3}. By decomposing the right-hand side in Eq.~\eqref{app:eq:tanhr3} in terms of squares of real and imaginary terms and then taking the square root, we find 
\begin{align} \label{app:eq:rewritten:tanh3}
\tanh(r_3) &= \sqrt{1-\frac{2}{\cosh (4 J_-) \cosh (4 J_+)+1}}. 
\end{align}
Then, using Eqs.~\eqref{app:eq:relate:P:Q:J}, which relate $J_\pm$ to  the functions $P(t)$ and $Q(t)$, and inverting Eq.~\eqref{app:eq:rewritten:tanh3} for $r_{3}$, which we rename to $r_{sq}$ as in the main text,  we find 
\begin{align} \label{app:eq:def:of:rsq}
   r_{3} \equiv  r_{sq} & =  \mathrm{arctanh} \sqrt{1 -\frac{4}{2 + P^2(t) + Q^2(t) + \dot{P}^2(t) + \dot{Q}^2(t)}}. 
\end{align}
Let us analyze this expression for $r_{sq}$. The initial conditions for $P(t)$ and $Q(t)$ read: $P(t =0) = 1$ and $Q(t=0)=1$. This implies that there is zero squeezing $r_{sq} = 0$ at $t = 0$, which is what we expect. Furthermore, since $P(t)$ and $Q(t)$ in Eq.~\eqref{app:eq:P:Q} grow exponentially with $t$, $r_{sq}$ tends to infinity in the limit of large  $t$. 

To summarize, we have shown that the parametric modulation imposes the unitary transformation of the state 
\begin{align} \label{app:eq:decomposed:phase:squeezing}
\hat U(t) =  e^{- i J_0 \hat a^\dag \hat a } \, e^{i \varphi_{sq} (\hat a^\dag \hat a + 1)} \, S(z_{sq}), 
\end{align}
where $z_{sq} = r_{sq} e^{i \theta_{sq}}$. The squeezing magnitude $r_{sq}$ is given in Eq.~\eqref{app:eq:def:of:rsq} and we find 
\begin{align} \label{app:eq:phisq}
e^{i \theta_{sq}} & = \frac{\cosh(r_{sq})}{\sinh(r_{sq})} \frac{i  \sinh(2 J_+ ) \cosh(2J_-) -  \cosh(2J_+) \sinh(2J_-)}{\cosh(2J_+) \cosh(2J_-) - i  \sinh(2J_+) \sinh(2J_-) }, 
\end{align}
the phase $\varphi_{sq}$ in Eq.~\eqref{app:eq:decomposed:phase:squeezing} is given by ~\cite{schneiter2020optimal}
\begin{align} \label{app:eq:args}
    \varphi_{sq} =  \rm{arg}\left[ \frac{1 - i \tanh(2J_+) \tanh(2J_-) }{1 + i  \tanh(2J_+) \tanh(2J_-) }\right].
\end{align}
If we ignore the global phase in Eq.~\eqref{app:eq:decomposed:phase:squeezing}, we can write \begin{align} \label{app:eq:decomposed:phase:squeezing}
\hat U(t) =  e^{- i \varphi(t) \hat a^\dag \hat a } \, S(z_{sq}), 
\end{align}
where $\varphi(t) = J_0 - \varphi_{sq}$. 

\section{Approximate solutions to the dynamics} \label{app:dynamics}
In this appendix we outline the derivation of the dynamics generated by the Hamiltonian in Eq.~\eqref{eq:main:Hamiltonian1}. We solve the dynamics perturbatively using two separate methods: first, we use a well established two-time perturbative solution of the Mathieu equation, which underpins the dynamics, and second, we use a perturbative solution of the unitary time-evolution operator $\hat U(t)$ in Eq.~\eqref{eq:time:evolution:operator} to first order. We also examine the stability and error of the solutions, where we show that the approximate solution for $\hat U(t) $ produces an error in the Bogoliubov coefficients that grows in time, while the Mathieu equation result in an error that oscillates in time.

\subsection{Approximate solutions to Mathieu’s equation} \label{app:mathieu}

Mathieu's equation can be approximately solved using a  standard two-time solutions, see e.g. Ref~\cite{kovacic2018mathieu}. The solutions were previously presented in Refs~\cite{qvarfort2020time} and~\cite{qvarfort2021optimal} and are valid for $2\lambda/\omega_0 \ll 1$. The approximate solutions for $P(t)$ and $Q(t)$ are, where we have rescaled $t\omega_0 \rightarrow t$ and $\lambda/\omega_0 \rightarrow \lambda$~\cite{qvarfort2020time, qvarfort2021optimal}:
\begin{equation}  \label{app:eq:P:Q}
\begin{split}
P(t) &=\frac{[ \lambda \cos(t + \phi_p) - \cos(t) ] \cosh( \lambda t) + [ \sin(t + \phi_p)- \lambda \sin(t) ] \sinh( \lambda t) }{ \lambda \cos(\phi_p) - 1} ,  \\
Q(t) &= \frac{\cos(t + \phi_p) \sinh(\lambda t) - \sin(t) \cosh(\lambda t)}{\lambda \cos(\phi_p) - 1}, 
\end{split}
\end{equation}
which are used to derive $\alpha(t)$ and $\beta(t)$ in Eq.~\eqref{eq:approximate:Bogoliubov1} after expanding in $\lambda$ to first order, and with factors of $\omega_0$ restored.

\subsection{Time-evolution perturbation theory} \label{app:perturbation:theory}
We now present an alternative method by which the dynamics can be solved. To treat the dynamics of the Hamiltonian in Eq.~\eqref{eq:main:Hamiltonian1}, we make use of time-dependent perturbation theory. We start by writing down the time-evolution operator $\hat U(t)$
\begin{align}
\hat U(t) = \overleftarrow{\mathcal{T}} \mathrm{exp} \left[ - \frac{i}{\hbar} \int^t_0 \mathrm{d} t' \, \hat H(t') \right]. 
\end{align}
We now divide the Hamiltonian in Eq.~\eqref{eq:main:Hamiltonian1} into two parts: one that contains a modified free evolution term with $\hat a^\dag \hat a $, and the other part that contains the interaction term
\begin{equation}
\begin{split}
\hat H_0(t) &= \hbar \left[ \omega_0 + 2 f(t) \right] \hat a^\dag \hat a ,  \\
\hat H_I(t) &= \hbar f(t) \left( \hat a^{\dag 2} + \hat a^2 \right),  
\end{split}
\end{equation}
where we have ignored a scalar term since it results in a global phase. 
We then consider the frame that rotates with $\hat H_0(t)$. For the choice of $f(t) = \lambda \cos(2\omega_0 t + \phi_p)$ in this paper, the  evolution generated by $\hat H_0(t)$ is given by 
\begin{align}
\hat U_0(t) = \mathrm{exp} \left[ - i  \int^t_0 \mathrm{d}t' \, \left( \omega_0 + 2 f(t')\right)\hat a^\dag \hat a \right] = e^{- i \theta(t) \hat a^\dag \hat a } . 
\end{align}
where we have defined $\theta(t) = \omega_0 t + 2 \frac{ \lambda }{\omega_0} \cos(\omega_0 t + \phi_p) \sin(\omega_0 t) $. The interaction Hamiltonian in this frame evolves with $\hat U_0(t)$ such that 
\begin{align}
\hat H_I'(t) = \hat U_0^\dag (t) \,\hat H_I(t) \, \hat U_0(t) = \hbar f(t) \left( e^{2 i \theta(t) } \hat a^{\dag 2} + e^{- 2 i \theta(t)} \hat a^2 \right), 
\end{align}
where we  have used the fact that $e^{ i x \hat a^\dag \hat a }\,  \hat a  \,e^{- i x \hat a^\dag \hat a} = e^{- i x } \, \hat a$. 
The evolution operator in the interaction frame is therefore
\begin{align}  \label{app:eq:interaction:evolution}
\hat U_I(t) = \overleftarrow{\mathcal{T}} \mathrm{exp} \left[ - i  \int^t_0 \mathrm{d}t' \, f(t') \, \left( e^{2 i \theta(t^\prime) } \hat a^{\dag 2} + e^{- 2 i \theta(t^\prime)} \hat a^2 \right) \right]. 
\end{align}
Returning to the laboratory frame, the full evolution can be written as $\hat U(t) = \hat U_0(t) \hat U_I(t)$.  When $\lambda t \ll 1$, we can expand the exponential in Eq.~\eqref{app:eq:interaction:evolution} to first order in $\lambda$ to find 
\begin{align} \label{app:eq:perturbed:U}
\hat U_I(t) \approx  1 - i \lambda \int^t_0 \mathrm{d}t' \, \cos(2 \omega_0 t' + \phi_p) \left( e^{2 i \theta(t^\prime) } \hat a^{\dag 2} + e^{- 2 i \theta(t^\prime)} \hat a^2 \right). 
\end{align}
We then examine the evolution of $\hat a$ and find 
\begin{align} \label{app:eq:evolution:a}
\hat a(t) &= \hat U^\dag(t) \, \hat a \, \hat U(t) \nonumber \\
&= \hat U_I^\dag(t) \, \hat U_0^\dag(t) \, \hat a \, \hat U_0(t) \, \hat U_I(t)  \\
&=e^{- i \theta(t) } \,  \hat U_I^\dag(t)  \, \hat a \, \hat U_I(t). \nonumber
\end{align}
Then, we insert the approximate form of $\hat U_I(t)$ shown in Eq.~\eqref{app:eq:perturbed:U} into Eq.~\eqref{app:eq:evolution:a} to find 
\begin{equation}
\begin{split}
\hat a(t) & \approx e^{- i \theta(t) }  \left\{ \hat a  +  i \lambda \int^t_0 \mathrm{d}t' \, \cos(2\omega_0 t' + \phi_p) \left[ \left( e^{2 i \theta(t') } \hat a^{\dag 2} + e^{- 2 i \theta(t')} \hat a^2 \right), \hat a \right] \right\}   \\
& \approx  e^{- i \theta(t) } \left[ \hat a  - 2 i \lambda \int^t_0 \mathrm{d}t' \, \cos(2 \omega_0 t' + \phi_p) \,e^{2 i \theta(t')} \, \hat a^\dag \right]  .
\end{split}
\end{equation}
Expanding and evaluating the integral, we may identify the Bogoliubov coefficients $\alpha(t)$ and $\beta(t)$ as per Eq.~\eqref{eq:bg1}. We find, expanding $\lambda$ to first order, 
\begin{equation}
\begin{split} \label{app:eq:perturbation:theory:alpha:beta}
\alpha(t) & \approx e^{-i \omega_0 t}\left[1 -2 i   \frac{\lambda}{\omega_0} \cos(\omega_0 t + \phi_p) \sin(\omega_0 t) \right], \\
\beta(t) &\approx   - i \frac{\lambda}{2\omega_0}  e^{-i ( \omega_0 t +  \phi_p)} \left[ 2 \omega_0 t +  e^{2 i (\omega_0 t + \phi_p)}\sin(2 t \omega_0) \right] . 
\end{split}
\end{equation}
As can be seen, both coefficients contain linear corrections of $\lambda$, but they are a bit different from those derived in the preceding section.

\begin{figure*}
\subfloat[ \label{fig:Delta:P}]{
  \includegraphics[width=0.45\linewidth]{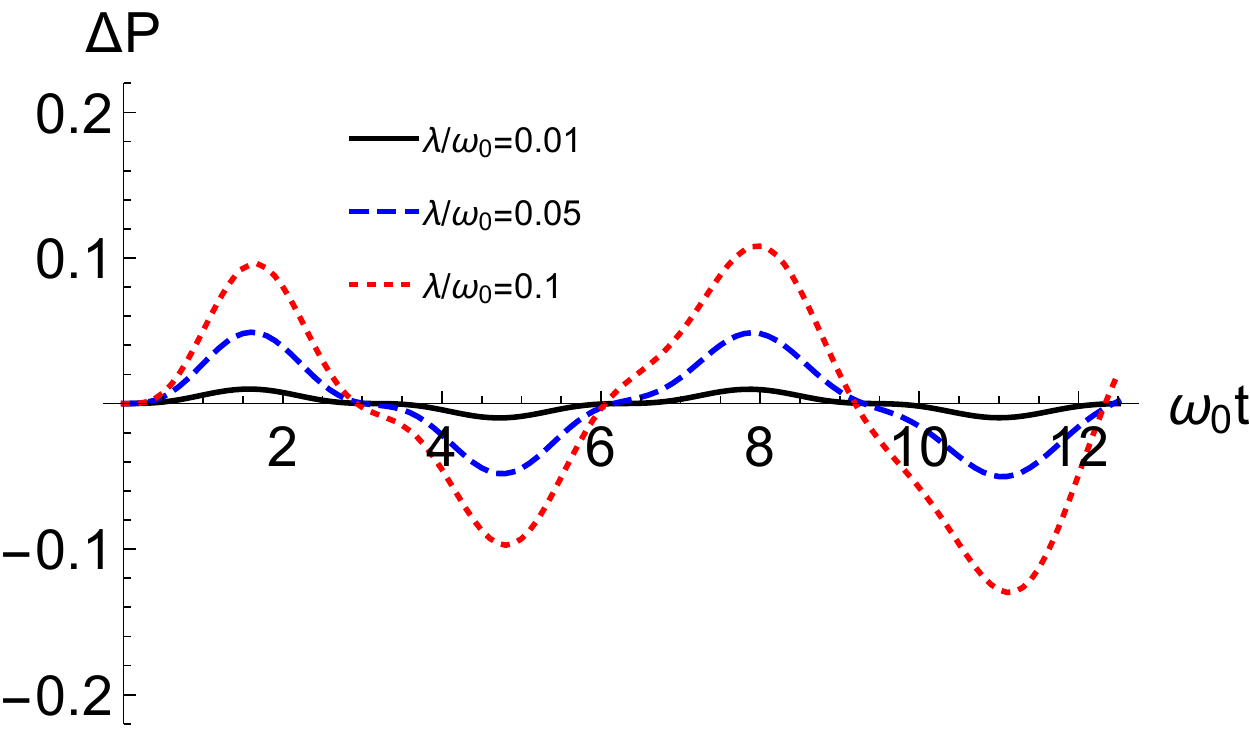}
}  
\subfloat[ \label{fig:Delta:Q}]{
  \includegraphics[width=0.45\linewidth]{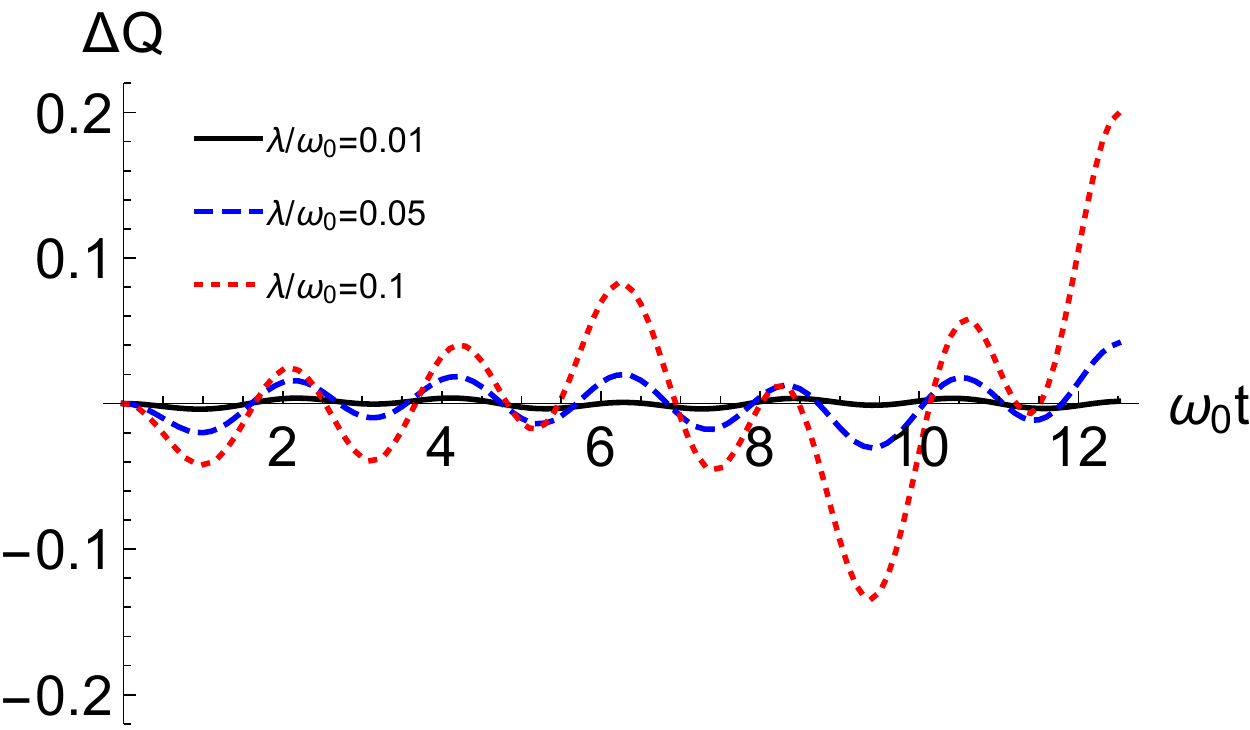}
}
\caption{ Plot showing the difference between a numerically obtained solution of Mathieu's equation and the approximate expressions in Eq.~\eqref{app:eq:P:Q}. Here $\Delta P$ and $\Delta Q$ are the differences between the numerically obtained solution and the approximate solutions plotted for various values of $\lambda/\omega_0$. As the driving strength increases, the errors start diverging. For the values used in this paper $\lambda/\omega_0 = 0.01$, the solutions stay stable and the errors small. }
\label{fig:P:Q:errors}
\end{figure*}

\subsection{Error analysis of the perturbative solutions} \label{app:sec:error:analysis}
The first step we perform in order to determine the error of the perturbative method in Appendix~\ref{app:mathieu} is to plot the solutions in Eq.~\eqref{app:eq:P:Q} against numerically obtained solutions of Mathieu's equation. We do so in Fig.~\ref{fig:P:Q:errors}, where we have defined $\Delta P(t)$ and $\Delta Q(t)$ as the deviations away from the numerical result. The phase is set to $\phi_p = \pi/2$. As can be seen, for short times the error remains similar in magnitude to the driving amplitude $\lambda/\omega_0$. 

Another way in which we can quantify the errors of our approximate solutions is by considering the Bogoliubov normalization relation $|\alpha|^2 - |\beta|^2 = 1$. Starting with the solutions obtained by expanding $\hat U(t)$, shown in Eq.~\eqref{app:eq:perturbation:theory:alpha:beta}, we find to second order in $\lambda$ that 
\begin{align} \label{app:eq:petrubative:bogoliubov}
|\alpha(t)|^2 - |\beta(t)|^2 &= 1 - \lambda^2 t^2   + \mbox{oscillating terms}. 
\end{align}
Here we note from Eq.~\eqref{app:eq:petrubative:bogoliubov} that the error grows with $t$, which means that the solutions derived in Appendix~\ref{app:perturbation:theory} will become increasingly inaccurate. 

In contrast, using the expressions for the Bogoliubov coefficients obtained from the perturbative solutions to the Mathieu equation,
we find that the error is given by 
\begin{equation} \label{app:eq:bogoliubov:error}
|\alpha(t)|^2 - |\beta(t)|^2 \approx  \frac{1 - \lambda \cos(2 t\omega_0 + \phi_p)/\omega_0}{1 - \lambda \cos(\phi_p)/\omega_0}. 
\end{equation}
We note that our solution is exact whenever $\phi_p = n  \pi/2$ and $2 \omega_0 t + \phi_p = n \pi/2$, for integer $n$. For example, when  $\phi_p = \pi/2$, the solution is exact at $\omega_0 t = \pi $. 

\begin{figure*}
\subfloat[ \label{fig:bogoliubov}]{
  \includegraphics[width=0.45\linewidth]{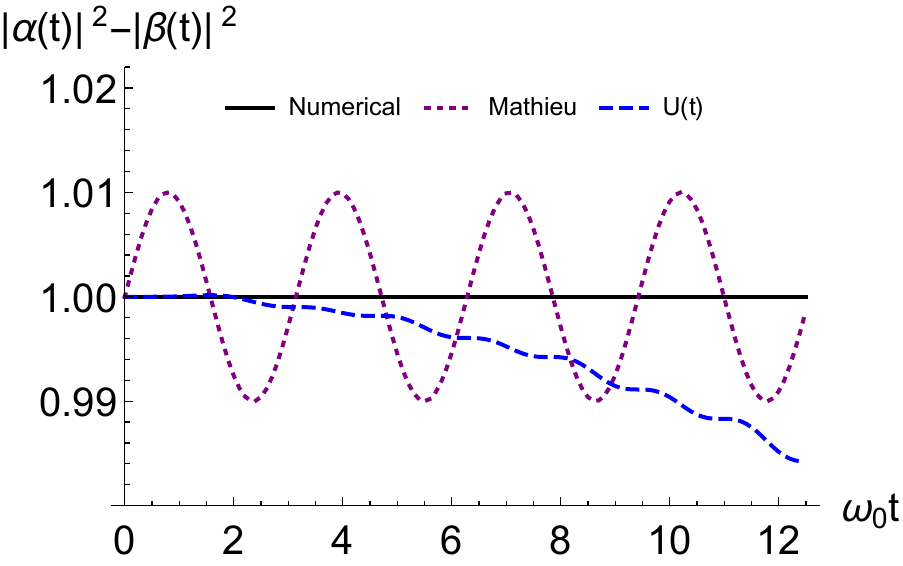}
}  
\subfloat[ \label{fig:quanta}]{
  \includegraphics[width=0.45\linewidth]{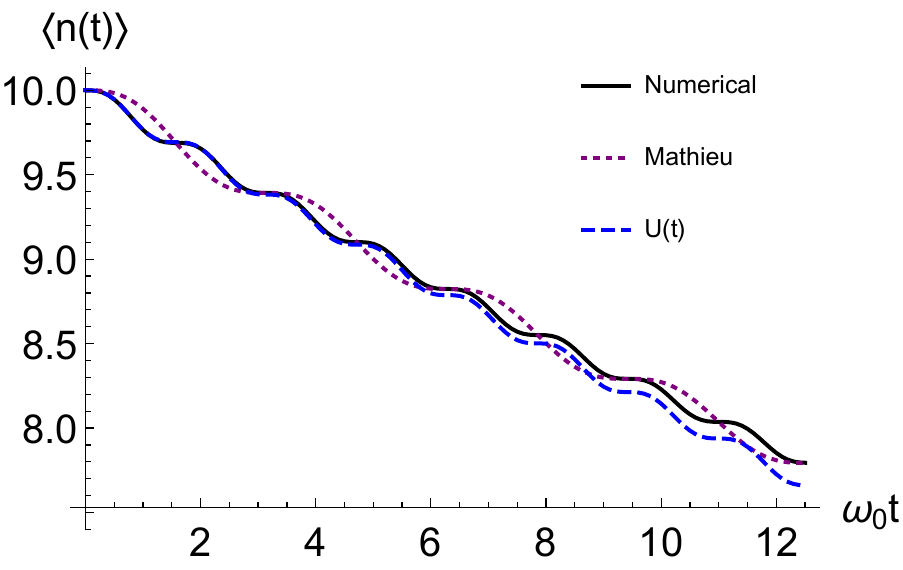}
}
\caption{ Plot showing the errors in the (a) Bogoliubov condition and (b) number of quanta given the numerically obtained solutions (black),  the perturbative solutions of Mathieu's equation in Eq.~\eqref{app:eq:P:Q} (purple dotted), and the solutions from the expanded $\hat U(t)$ in Eq.~\eqref{app:eq:perturbation:theory:alpha:beta} (blue dashed). The parameters are $\lambda/\omega_0 = 0.02$, $\phi = 0$, $\phi_p = \pi/2$, and $r = \sqrt{10}$. The approximate solution to Mathieu's equation follows the outline of the exact solution, but it fails to replicate some of the faster oscillations. As shown in Eq.~\eqref{app:eq:bogoliubov:error}, the approximate solution is accurate whenever $\omega_0t$ is a multiple of $\pi$, given that $\phi_p = \pi/2$.  }
\label{fig:Bogoliubov:quanta}
\end{figure*}

In Fig.~\ref{fig:Bogoliubov:quanta} we compare the error of the approximate solutions shown in Eqs.~\eqref{app:eq:perturbation:theory:alpha:beta} and ~\eqref{app:eq:P:Q} with a numerically obtained solution of Mathieu’s equation for an initial coherent state $\ket{r e^{i \phi}}$. The parameters are set to $\phi = 0$, $\phi_p = \pi/2$, $\lambda/\omega_0 = 0.01$ and $r = \sqrt{10}$. We note a few things from this figure. First, we note that the exact solutions (black solid lines) have a periodicity that is about twice that of the approximate solutions to Mathieu's equation (purple dotted line). The missed oscillations can also be observed as errors in Fig.~\ref{fig:P:Q:errors}. It might be possible to further improve the accuracy of the two-time scale solutions by adding a third time-scale, which is stretched by $\sqrt{\epsilon}$. We leave such an analysis to future work. 
Second, we note that the error of the solutions from expanding $\hat U(t)$ (blue dashed line) grows in time, and thereby diverges from the numerical solutions to a greater extent than those obtained from Mathieu's equation. They are however more accurate for shorter time-scales, since they reproduce the shorter oscillations of the numerically obtained solutions.

\subsection{Mathieu equation stability analysis}

The Mathieu equation is numerically unstable, which means that  certain parameter combinations result in diverging solutions~\cite{kovacic2018mathieu}. When $a = 1$, there are in fact no stable solutions that can be obtained. However, since we measure the state at the beginning of each cooling cycle, we effectively reset the instabilities that would have been introduced for the full running time of the protocol. In this way, the inclusion of measurements also prevents the buildup of instability from the modulations of the potential (See Fig. 4).

The  analytic extension to non-unitary dynamics is likely to change the stability of the equations of motion, however such a stability analysis would require a full analytical solution of the open systems dynamics with the time-dependent frequency modulation. We leave this to future work.

\section{Applying parametric modulations to a thermal state\label{thermalNote}}
Driving alone is insufficient to cool down arbitrary quantum states lacking a phase reference such as thermal states. To see this, we now consider the effect of applying the cycle to a thermal state of the quantum oscillator at inverse temperature $\beta' =1/k_{B}\text{T}$, given by
\begin{equation} \label{eq:thermal:state}
\hat \varrho_{\mathrm{th}} = \frac{e^{-\beta'\hat H_0}}{\mathcal{Z}},~\text{where}~ \mathcal{Z}=\text{tr}\{e^{-\beta'\hat H_0}\}.
\end{equation}
We can compute $ \braket{\hat n(t)}_{\mathrm{th}} $ to find, 
\begin{equation}
\begin{split}
 \label{eq:thermal:state:N}
\braket{\hat n (t) }_{\mathrm{th}} &= ( |\alpha(t)|^2  + |\beta(t)|^2) \bar{n} + |\beta(t)|^2  \\
&=  \bar{n} + (1 + 2 \bar{n}) |\beta(t)|^2,
\end{split}
\end{equation}
where we used the Bogoliubov identity $|\alpha(t)|^2 = |\beta(t)|^2 + 1$ and the fact that for thermal states, $\braket{\hat a^2(t)}_{\mathrm{th}} = \braket{\hat a^{\dag 2}(t)}_{\mathrm{th}} = 0$. Here $\bar{n} = \text{Tr}\{\hat \varrho_{\mathrm{th}} \, \hat n\}= (e^{\frac{\hbar\omega_{0}}{k_{B}\text{T}}}-1)^{-1}$. Since all quantities in Eq.~\eqref{eq:thermal:state:N} are positive, the mean quanta cannot decrease by the driving alone. We note that this is true \textit{regardless} of what dynamics we are considering, since this expression is completely general in terms of the Bogoliubov coefficients. In other contexts, the limits of algorithmic cooling with Gaussian resources have been considered~\cite{serafini2020gaussian}. 

\subsection{Feedback cooling a thermal state} We now examine the resulting average cooling for a single cycle given an initial thermal state measured in the coherent state basis. The probability of obtaining a specific coherent state $|\xi\rangle=|re^{i\phi}\rangle$ by performing a heterodyne measurement on a thermal state is given by the corresponding Husimi $Q$-function~\cite{husimi1940some}: $Q(\xi)=Q(r,\phi)=\frac{1}{\pi(\bar{n}+1)}e^{-\frac{r^{2}}{\bar{n}+1}}$. Here $\bar{n} = \text{Tr}\{\hat \varrho_{\mathrm{th}} \, \hat n\}= (e^{\frac{\hbar\omega_{0}}{k_{B}\text{T}}}-1)^{-1}$.
By then averaging over all possible outcomes of $r$ and $\phi$, we can determine the average cooling power. 

We find to first order in $(\lambda/\omega_0)$ that the occupation number on average is $\braket{ n (t)} = 1 + \bar{n}( 1 - 2 \lambda t) + \mathcal{O}[(\lambda/\omega_0)^2]$,  where the angular brackets here indicate averaging over many measurement outcomes. This means that the average cooling power at early times is $2\lambda \bar{n}$. 
 In Fig.~2(a) of the main-text, we compare the analytical prediction for the average quanta $\braket{n (t)}$  for a single cooling cycle with numerical simulations and found excellent agreement.

\section{Optimal cooling} \label{app:squeezing}

Here we derive the conditions for optimal cooling through parametric modulations and phase-preserving quantum measurements. 
From our derivation of the time-evolution operator $\hat U(t)$ in Appendix~\ref{app:relations}, we know that the application of parametric modulation corresponds to a rotation and two consecutive single-mode squeezing operations. By studying the total resulting squeezing, it is possible to determine for how long the protocol should be applied for in order to optimally cool the state towards its quantum ground-state.

\begin{figure*}
\includegraphics[width=0.45\linewidth]{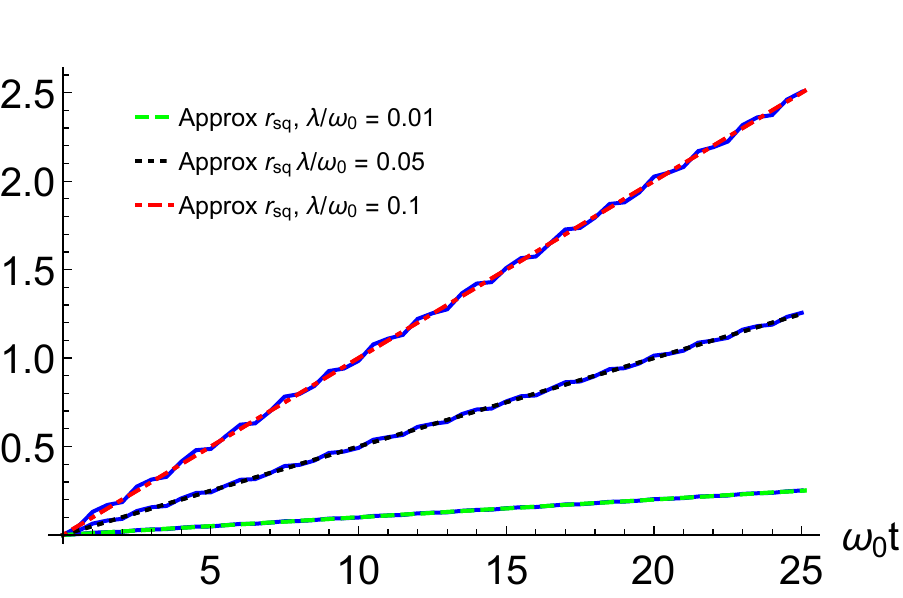}
\caption{Plot of the squeezing value $r_{sq}$ as a function of time $\omega_0 t$ for various values of the modulation strength $\lambda$. The dotted and dashed lines show the approximate value $r_{sq} \approx \lambda t$, while the solid lines show the numerically values for $r_{sq}$, which have been obtained by numerically solving Mathieu’s equation. The approximate values follow the leading-order linear behavior of the numerically obtained values. }
\label{fig:r:plot}
\end{figure*}

To determine the leading-order behavior of $r_{sq}$, we expand the function under the square root in Eq.~\eqref{app:eq:def:of:rsq} for small driving strength $\lambda$. Using the approximate expressions for $P(t)$ and $Q(t)$ in Eq.~\eqref{app:eq:P:Q} (where we have redefined time $t$ and $\lambda$ in units of $\omega_0$), we find 
\begin{align} \label{app:eq:approximated:rsq:expansion}
[\mathrm{arctanh}^{-1}(r_{sq})]^2 = 1 -\frac{4}{2 + P^2(t) + Q^2(t) + \dot{P}^2(t) + \dot{Q}^2(t)} \approx \frac{1}{8}   \left[\lambda^2  \left(8 t^2+\cos (4 t)-1\right)+4 \lambda \sin (2 t)\right]. 
\end{align}
We then focus on the term $\lambda^2t^2$, which grows quadratically in time, and ignore the oscillating terms, since they just create perturbations around this value. Taking the square root, we are left with just $\lambda t$. Then, we note that as long as $\lambda t$ remains small, we can use the expansion for $\mathrm{arctanh}(x) $, which reads $\mathrm{arctanh}(x) \approx z + z^3 /3 + \ldots $. Thus we find the surprisingly simple linear scaling for the total squeezing, namely
\begin{align} \label{app:eq:approx:rsq}
r_{sq} \approx  \lambda t. 
\end{align}
We plot this expression in Eq.~\eqref{app:eq:approx:rsq} alongside the numerically obtained value for $r_{sq}$ as a function of time in Fig.~\ref{fig:r:plot} for different values of the modulation strength $\lambda/\omega_0$. The  blue solid lines show $r_{sq}$ for $\lambda/\omega_0 = 0.01, 0.05$ and $0.1$ respectively. The dashed and dotted lines show the approximate value  $\lambda t$. We note that the approximate expression in Eq.~\eqref{app:eq:approx:rsq} fully captures the leading-order linear behavior of $r_{sq}$.

The question now becomes what the optimal modulation time is. We can answer this question by studying the occupation number $\braket{\hat a^\dag \hat a}$ after applying the combined squeezing operator in Eq.~\eqref{app:eq:decomposed:phase:squeezing}. This allows us to determine the optimal value for $r_{sq}$, which in turn tells us for how long the modulations should be turned on. 

We start by computing the photon number for the effective squeezing operator $\hat S(z_{sq})$ in Eq.~\eqref{app:eq:decomposed:phase:squeezing}. For an initial coherent state $\ket{\xi}$, with $\xi = r e^{i \phi}$ as in the main text, the number of quanta are given by 
\begin{align} \label{app:eq:squeezing:quanta}
 \bra{\xi} \hat S^\dag(z_{sq}) \, \hat a^ \dag \hat a \, \hat S(z_{sq}) \ket{\xi} =
r^2 \cosh^2(r_{sq}) -2 r^2 \cos(2 \phi - \phi_{sq}) \cosh(r_{sq}) \sinh(r_{sq}) + (r^2 + 1) \sinh^2(r_{sq}) . 
\end{align}
From studying Eq.~\eqref{app:eq:squeezing:quanta}, we see that the squeezing operation reduces the number of quanta provided that $2\phi - \phi_{sq} = 2 \pi n $, where $n$ is an integer. Note that this phase relation is different from that in the main text, because $\phi_{sq}$ in Eq.~\eqref{app:eq:phisq} is non-trivially related to the parametric modulation phase $\phi_{p}$. 

We now note that if $r_{sq}$ is too large, the system gains quanta instead. For each coherent state occupation number $r$, there exists an ideal squeezing value  $r_{sq}$ which minimizes $\bra{\xi} \hat S^\dag(z_{sq}) \, \hat a^ \dag \hat a \, \hat S(z_{sq}) \ket{\xi}$ for a particular $ r_{sq}$. To find this  $r_{sq}$, we differentiate Eq.~\eqref{app:eq:squeezing:quanta} with respect to  $r_{sq}$ to find
\begin{align} \label{app:eq:diff:squeezing}
\frac{d}{dr_{sq}}\bra{\xi} \hat S^\dag(z_{sq}) \, \hat a^ \dag \hat a \, \hat S(z_{sq}) \ket{\xi} = \sinh(2r_{sq}) + 2 r^2 \left[ \sinh(2 r_{sq}) - \cos( 2 \phi - \phi_{sq}) \cosh(2r_{sq}) \right]. 
\end{align}
Setting Eq.~\eqref{app:eq:diff:squeezing} to zero and solving for $r_{sq}$ using the optimal phase relation $2 \phi - \phi_{\color{blue} sq} = 0$, we find that the optimal squeezing value for a specific value of $r$ is given by 
\begin{equation} \label{app:eq:optimal:rsq}
r_{sq}= \frac{1}{4}\mathrm{log}(1 + 4r^2).
\end{equation}
Inserting this result back into Eq.~\eqref{app:eq:squeezing:quanta}, we obtain 
\begin{equation}
\bra{\xi} \hat S^\dag(z_{sq}) \, \hat a^ \dag \hat a \, \hat S(z_{sq}) \ket{\xi} = \frac{1}{2} (\sqrt{1 + 4 r^2} - 1), 
\end{equation} 
which is the lowest number of quanta a coherent state with coherent state magnitude $r$ can be cooled to.  If we squeeze beyond this value, quanta are added to the system rather than removed. For example, given a coherent state with $r = 1$, the optimal squeezing value is $r_{sq} =  \mathrm{log}(5)/{\color{blue}4}$, which results in $\bra{\xi}\hat S^\dag(r_{sq}) \hat a^ \dag \hat a \hat S(r_{sq}) \ket{\xi} \approx 0.6$. It is not possible to reduce the number of quanta beyond this value by squeezing alone. 

By knowing the optimal squeezing value and the approximate expression for $r_{sq} \approx \lambda t$, it is possible to derive the optimal modulation time. We know from Eq.~\eqref{app:eq:approx:rsq} that $r_{sq} \approx \lambda t$. By equating this to the optimal squeezing value and solving for time, we find the optimal modulation time $t_{op}\approx \log(1 + 4 r^2)/4\lambda$. For an initial coherent state with $r = 1$ and a modulation strength $\lambda/\omega_0 = 0.01$, the optimal modulation time is $t_{op} \approx 40/\omega_0$. We note that for low occupation number $r$, the optimal modulation time is short, which might make it challenging to cool the state optimally. 

It should be noted that these results only apply to closed-system dynamics. In the presence of finite thermal dissipation, the optimal time occurs earlier than that predicted here. In Fig. 2(c), we numerically computed the occupation value for dissipative dynamics and determine at what time the minimum value is achieved. We then terminated the modulation protocol at the closest multiple of $\pi$, which is where the potential returns to its original value. To analytically determine the optimal modulation time for dissipative dynamics, one would have to solve the master equation analytically. We leave this to future work.

\section{Protocol with strong homodyne measurements} \label{app:homodyne:measurements}
\begin{figure*}
\includegraphics[width=0.45\linewidth]{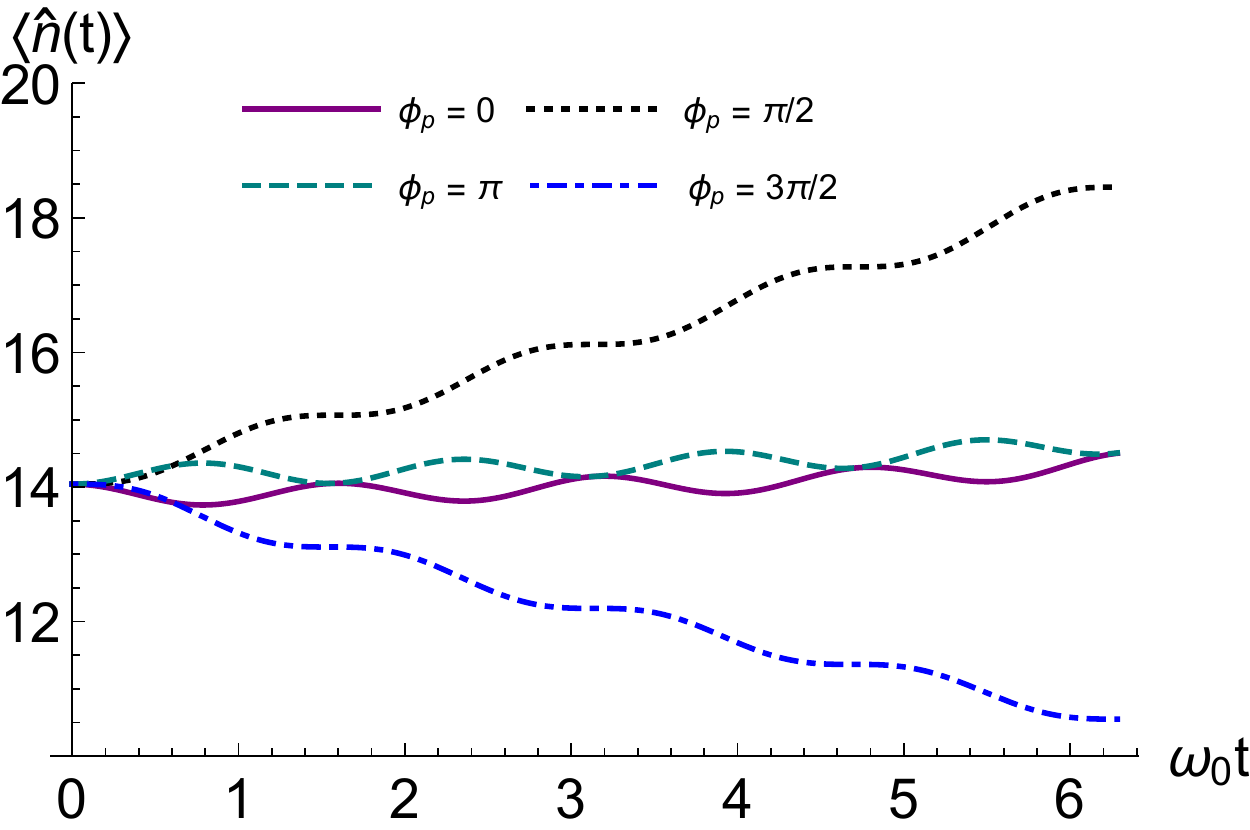}
\caption{Plot of the number of quanta $\braket{\hat n(t)} = \braket{\hat a^\dag \hat a }$ given an initial position eigenstate $\ket{x_\theta}$. The parameters are $x = 0.5$, $\theta = 0$, and $\lambda /\omega_0 = 0.02$. We observe cooling when the trapping potential is parametrically modulated with the phase-offset $\phi_p = 3\pi/2$. }
\label{fig:homodyne:plot}
\end{figure*}

Instead of projecting into the coherent-state basis, we may also consider homodyne measurements of the state. For completeness, we show here that the quanta of position eigenstates can also be reduced using the parametric modulations, however we find that the phase relation differs from that identified in the main text for coherent states. 

We start again with the fact that $\hat a(t) = \alpha(t) \hat a + \beta(t) \hat a^\dag$. 
We then let  the initial state $\ket{x_\theta}$ be an eigenstate of the generalized quadrature operator $
\hat x_\theta = (e^{i \theta} \hat a^\dag + e^{- i \theta} \hat a )/\sqrt{2}$, where $\theta$ is a phase. 
We can then define eigenstates of $\hat x_\theta$ such that $\hat x_\theta \ket{x_\theta} = x_\theta \ket{x_\theta}$, where $x_\theta$ is the eigenvalue. Note that $\ket{x_\theta}$ is not a proper normalized eigenstate because it is not square-integrable.  The overlap with the Fock state $\ket{n}$ is however well-defined
\begin{eqnarray}
\braket{\hat x_\theta| n} = \frac{1}{\pi^{1/4} \sqrt{2^n n!}}  e^{- x^2/2} H_n(x)\, e^{- i n \theta}, 
\end{eqnarray}
where $H_n(x)$ is the Hermite polynomial. This means that we can expand the position eigenstates in the Fock basis as
\begin{align} \label{app:eq:position:eig:expansion}
\ket{x_\theta} = \sum_{n=0}^\infty \ket{n}\braket{n|x_\theta} = \sum_{n=0}^\infty \frac{1}{\pi^{1/4} \sqrt{2^n n!}}  e^{- x^2/2} H_n(x) \, e^{i n \theta} \ket{n}. 
\end{align} 
The number of quanta for a single position eigenstate is then given by 
\begin{align} \label{eq:homodyne:exp:value}
\braket{\hat a^\dag \hat a(t)}_{\ket{x_\theta}} &= |\alpha(t)|^2\bra{x_\theta} \hat a^\dag \hat a \ket{x_\theta} + \alpha^*(t) \beta(t)\bra{x_\theta}  \hat a^{\dag 2} \ket{x_\theta} + \beta^*(t) \alpha(t)\bra{x_\theta} \hat a^2 \ket{x_\theta} + |\beta(t)|^2 (\bra{x_\theta} \hat a^\dag \hat a\ket{x_\theta} + 1 ). 
\end{align}
We then use the Fock basis expansion in Eq.~\eqref{app:eq:position:eig:expansion} to find
\begin{align} \label{eq:expectation:values:hermite:polynomials}
\bra{x_\theta} \hat a^\dag \hat a \ket{x_\theta} &= \sum_{n=0}^\infty\frac{n}{\pi^{1/2} 2^n n!}  e^{- x^2} H_n^2(x),   \nonumber \\
\bra{x_\theta}\hat a^{\dag 2} \ket{x_\theta} &=\sum_{n=0}^\infty  \frac{1}{\pi^{1/2} 2^{n+1} n!} e^{-x^2} H_{n+2}(x) H_{n}(x) \, e^{- 2 i \theta},   \\
\bra{x_\theta} \hat a^2 \ket{x_\theta} &= \sum_{n=0}^\infty  \frac{1}{\pi^{1/2} 2^{n+1} n!} e^{-x^2} H_n(x) H_{n+2}(x) \, e^{ 2 i \theta}. \nonumber
\end{align}
Here, we note that the expressions on the last two lines in Eq.~\eqref{eq:expectation:values:hermite:polynomials} can be negative, since the Hermite polynomials contain odd powers of $x$ for odd $n$. Thus it is possible to contain cooling with a strong homodyne measurement as well. 

We now wish to derive an expression for the number of quanta akin to that in Eq.~\eqref{eq:coherent:state:cooling1}, which shows the occupation number for coherent states. Such an expression tells us what phase relationship we need between the phase of the generalized quadrature eigenstate $\theta$ and the phase of the parametric modulation $\phi_p$ for the system to be cooled. Since only the second two lines in Eq.~\eqref{eq:expectation:values:hermite:polynomials} have phases, we write them as $\bra{x_\theta} \hat a^{\dag 2} \ket{x_\theta} = e^{- 2 i \theta}  \bar{x}_\theta$, where 
\begin{align}
\bar{x}_\theta = \sum_{n=0}^\infty  \frac{1}{\pi^{1/2} 2^{n+1} n!} e^{-x^2} H_n(x) H_{n+2}(x). 
\end{align}
Inserting this into Eq.~\eqref{eq:homodyne:exp:value} and expanding to first order in $\lambda$ (which requires us to assume that $x$ is small), we find 
\begin{align} \label{eq:position:eigenstate:cooling}
\braket{\hat a^\dag \hat a (t)}_{\ket{x_\theta}} = \bra{x_\theta} \hat a^\dag \hat a \ket{x_\theta}  + \frac{\lambda}{\omega_0} \left\{  \bra{x_\theta} \hat a^\dag \hat a \ket{x_\theta} [\cos(\phi_p) -  \cos(2 \omega_0 t + \phi_p)] - 2  t \omega_0 \, \bar{x}_\theta \sin(2 \theta + \phi_p) \right\} + \mathcal{O}(\lambda^2). 
\end{align}
The number of quanta can only decrease if the last term of Eq.~\eqref{eq:position:eigenstate:cooling} is negative. To find out whether that is the case, we must first examine the sign of $\bar{x}_\theta$. Around $x\sim 0$, the following recurrence relation for the Hermite polynomials holds: $H_{n+2}(0) = - 2 (n+ 1) H_{n}(0)$. Using this, we find that 
\begin{align} \label{app:eq:bar:x:sum}
\bar{x}_\theta (x \sim 0 ) = - 2 \sum_{n=0}^\infty \frac{(n+1)}{\pi^{1/2} 2^{n+1} n!} H_n^2(0). 
\end{align} 
Since all terms inside the sum in Eq.~\eqref{app:eq:bar:x:sum} are positive, we deduce that $\bar{x}_\theta < 0$ for small $x$. This means that we require the term with $\sin(2\theta + \phi_p)$ to be maximally negative, which is true when $2\theta + \phi_p = 3\pi/2$. We note that this phase relationship is different from the coherent states, which required $2\phi + \phi_p = \pi/2$.  We plot $\braket{\hat n(t)}_{\ket{x_\theta}}$ as a function of time in Fig.~\ref{fig:homodyne:plot} for various choices of the phase $\phi_p$. The parameters are $x = 0.5$, $\theta = 0$, and $\lambda/\omega_0 = 0.02$. As expected, for this value of $x$, the system is cooled when $\phi_p = 3\pi/2$. 

For larger $x$, however, $\bar{x}_\theta$ in Eq.~\eqref{app:eq:bar:x:sum} is positive (this can be checked numerically), which means that the system is instead cooled when $2 \theta + \phi_p = \pi/2$. Since we cannot deterministically prepare the system in the eigenstate $\ket{x_\theta}$ where $x_\theta$ is small, we conclude that while homodyne measurements are an option for feedback cooling, heterodyne measurements are more reliable since the phase relation for coherent states remains the same regardless of the measurement outcome.

\section{Analysis of an ensemble of quantum trajectories} \label{app:trajectories}

It is possible to consider the quantum state as it goes through a number of measurements and parametric modulations that squeeze the coherent state. Each measurement has a particular probability $P(\xi_{j} |r^{j}_{sq}, \xi_{j-1})$ for returning the coherent state $\xi_{j}$ given the initial state $\xi_{j-1}$, determined by the Husimi $Q$ function. The probability also depends on $r^{j}_{sq}$, which results from the parametric modulations. 

For example, let the initial state after a projective measurement in the coherent state basis be $\ket{\xi_0}$, where we choose $\xi_0$ to be real, which can be achieved by applying a rotation after the measurement. The final state after the evolution is $\ket{\psi(t) } = \hat S(r_{sq}) \ket{\xi_0}$. The value of $r_{sq}$ is always set to its optimal value, which depends on the coherent state parameter $\xi$, as $r_{sq} =\log(1+4|\xi|^2)/4 $. 

Measurements in the coherent state basis are modeled by Kraus operators, $\frac{1}{\sqrt{\pi}}|\xi\rangle\langle\xi|.$ The corresponding prorability distribution of measurement outcomes is~\cite{albano2002squeezed},
\begin{equation}
\begin{split}
Q(\xi_{1},\xi_{0},r_{sq}) &= \frac{1}{\pi}|\bra{\xi_{1}} \hat S(z) \ket{\xi_{0}}|^{2}  \\
&=  \frac{1}{\pi\cosh(r_{sq})} \, \mathrm{exp} \left[ -  |\xi_{1}|^2  - |\xi_{0}|^2 - \xi_{1}^{*2} \tanh(r_{sq})  + \xi_{0}^2\tanh(r_{sq})   + \xi_{0} \xi_{1}^* /  \cosh(r_{sq})  \right]. 
\end{split}
\end{equation}
The probability of a sequence of measurement outcomes $\{\xi_{j}\}$ intervened by parametric modulations
is given by,
\begin{equation}
    P(\xi_{j},\xi_{j-1},...\xi_{0}) = \prod_{j=0}^{j-1}Q(\xi_{j},|\xi_{j-1}|,r^{j-1}_{sq}).
\end{equation}
Note the appearance of $|\xi_{j-1}|$ in the above expression, which accounts for the fact that we also rotate the coherent state to the real axis after each measurement (prior to parametric modulations).

\begin{figure*}
    \centering
   \includegraphics[width=0.5\linewidth]{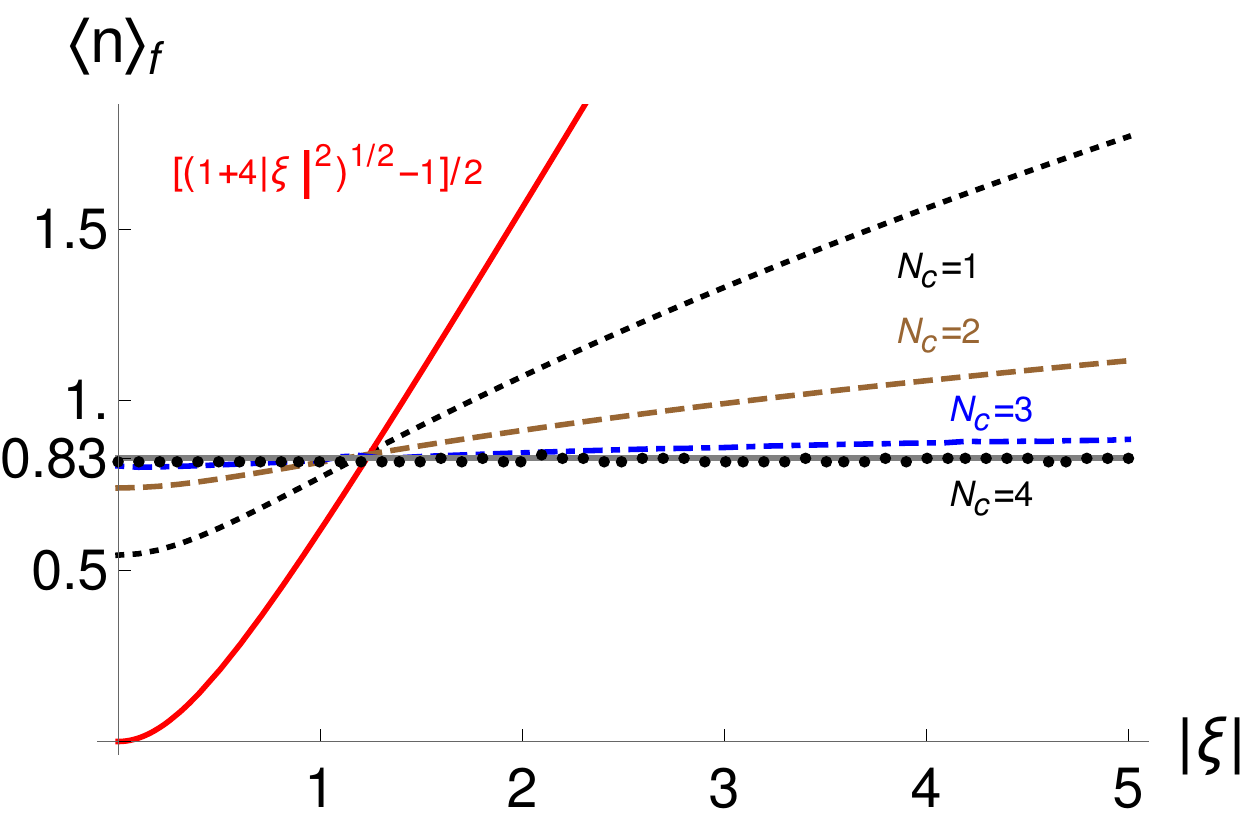}
    \caption{Plot showing the solution of Eq.~\eqref{app:eq:Fredholm}. Here, by numerically integrating the exact $Q$ function, we plot the final occupation number  $\langle n\rangle_{f}$ as a function of the initial coherent state amplitude $|\xi|$ for simply modulating for the optimal duration (red solid line), subsequently measured and modulated (one cycle, $N_{c}=1$)(black dotted line), and similarly up to four cycles. We see that the final occupation number achieved in our protocol for a larger cycle becomes independent of the initial coherent state parameter $\xi$ and achieves the value $\langle n\rangle_{f} = 0.83$. It also corresponds to an invariant cycle of cooling.}
    \label{fig:my_label}
\end{figure*}

The final average occupation number arrived at in the end, after averaging over all possible measurement outcomes is given by,
\begin{equation}
 \langle n\rangle_{j} =    \int d^{2}\xi_{0}
d^{2}\xi_{1}...d^{2}\xi_{j}P(\xi_{j},\xi_{j-1},...\xi_{0})\frac{1}{2} \bigg( \sqrt{ 1 + 4 |\xi_{j}|^2} - 1\bigg).
\end{equation}
By now requiring that $\langle n_{j}\rangle = \langle n_{j-1}\rangle$, we can derive the minimum value of $\langle n\rangle$ that appears to be a steady state value in Fig.~2(c). In fact, by requiring this, we arrive at the condition for an invariant cycle for mean quanta starting in a coherent state $\xi_{j-1}$:
\begin{equation} \label{app:eq:Fredholm}
    \langle n\rangle_{f}  = \frac{1}{2} \bigg( \sqrt{ 1 + 4 |\xi_{j-1}|^2} - 1\bigg)=\int d^{2}\xi_{j} Q[\xi_{j},|\xi_{j-1}|,r_{sq}(\xi_{j-1})]\bigg( \sqrt{ 1 + 4 |\xi_{j}|^2} - 1\bigg)/2.
\end{equation}
We solve this equation for consistency via numerical integration, which shows that a unique solution exist for $\xi$ such that $\langle n\rangle_{f}\approx 0.83$. See Fig.~\ref{fig:my_label}.

\end{document}